\documentclass[12pt]{iopart}
\usepackage{iopams}
\usepackage{graphicx}

\begin{document}

\title[Spherical spaces]{CMB radiation in an inhomogeneous spherical space}
\author{R.\,Aurich$^1$, P.\,Kramer$^2$ and S.\,Lustig$^1$}

\address{$^1$Institut f\"ur Theoretische Physik, Universit\"at Ulm,\\
Albert-Einstein-Allee 11, D-89069 Ulm, Germany}
\address{$^2$Institut f\"ur Theoretische Physik der Universit\"at,\\
Auf der Morgenstelle 14, D-74076 T\"ubingen, Germany}

\begin{abstract}
We analyse the CMB radiation in spherical 3-spaces
with non-trivial topology.
The focus is put on an inhomogeneous space which possesses observer
dependent CMB properties.
The suppression of the CMB anisotropies on large angular scales is
analysed with respect to the position of the CMB observer.
The equivalence of a lens space with a Platonic cubic space is shown
and used for the harmonic analysis.
We give  the transformation of the CMB multipole radiation amplitude
as a function of the position of the observer. General sum rules are obtained
in terms of the squares of the expansion coefficients for invariant polynomials
on the 3-sphere.
\end{abstract}


\submitto{\PS}

\section{Introduction.}
\label{sec:intro}

Cosmic topology examines multi-connected manifolds as candidates for
the spatial part of cosmic space-time. 
It simulates the cosmic microwave background (CMB) radiation using the 
eigenmodes of the multi-connected manifolds,
and explores if the specific multipole amplitudes and
selection rules arising from the manifold are encoded in the CMB radiation.

In this paper the focus is put on spherical spaces.
The simply connected 3-sphere underlies Einstein's initial
cosmological analysis of 1917 \cite{Einstein_1917}.
In that year de Sitter \cite{deSitter_1917}
already discussed the projective space $\mathbb{P}^3$
as an alternative to Einsteins model.
This was the first cosmological application of a spherical model
which is multiply connected,
i.\,e.\ being a closed three-dimensional piece of Einstein's 3-sphere.

The topology of a manifold ${\cal M}$ is locally described by homotopy. 
Homotopy composes loops on the manifold by concatenation and explores the 
homotopy group $\pi_1({\cal M})$ formed by these loops.
If any loop can be continuously contracted to a point, 
the manifold is simply connected.
Any topological manifold has an image on a simply connected covering manifold. 
The covering manifold $\tilde{{\cal M}}$ is tiled by copies of ${\cal M}$.
Covering manifolds and their tiles can be spherical, Euclidean or hyperbolic. 
In terms of the Riemannian metric they display constant positive,
zero or negative curvature.
In this paper, we concentrate on spherical manifolds
which tile the 3-sphere ${\mathbb S}^3$.

The second concept of a cover offers a global and
quasi-crystallographic view of a topological manifold.  
On the covering manifold $\tilde{{\cal M}}$ there is a group
${\rm deck}({\cal M})$ of deck transformations
which by fix-point free action tiles the covering manifold.
Seifert and Threlfall \cite{SE34} show that the two groups are isomorphic, 
$H={\rm deck}({\cal M})\sim\pi_1({\cal M})$, and so
the first local and the second global view of topology,
in terms of the group $H$ as topological invariant,
are equivalent.

In an abstract approach, a topological manifold is taken as a quotient space 
$\tilde{{\cal M}}/H$ of the covering manifold $\tilde{{\cal M}}$
by the group $H$.
If this notion refers to the abstract group and not to a representation thereof,
it leaves open the geometric form of the manifold.
An algebraic characterisation for  the classification of spherical space forms is
given by Wolf \cite{WO84} in terms of unitary matrix representations 
of groups $H$, acting on the 3-sphere.

An introduction into the topological concept applied to the cosmological
framework can be found in \cite{LachiezeRey_Luminet_1995,Levin_2002}
where all three spatial curvatures are discussed.
The focus was shifted to spherical spaces \cite{Gausmann_et_al_2001}
by the paper \cite{LU03}
which claims that the low power in CMB anisotropies at large scales
can be described by the Poincar\'e dodecahedral topology.
Thereafter, a lot of papers discussed the relevance of this result,
and other spherical spaces such as the truncated cube and the tetrahedral space
were investigated with respect to
their statistical CMB properties, see e.\,g.\
\cite{Roukema_et_al_2004,Gundermann_2005,AU05,AU05b,AU06,%
Lustig_2007,Roukema_Kazimierczak_2011}.
In addition, spherical lens spaces $L(p,q)$ are studied
in \cite{Uzan_et_al_2003}.
The fundamental domain can be visualised by a lens-shaped solid
where the two lens surfaces are identified by a $2\pi q/p$ rotation
for relatively prime integers $p$ and $q$ with $0<q<p$.
For more restrictions on $p$ and $q$, see  below and \cite{Gausmann_et_al_2001}.
A further family of spherical spaces,
the so-called Platonic polyhedra, were constructed from their homotopy groups
and studied in \cite{KR05,KR10,KR10B,KR10C}.

In the present paper, we examine the equivalence of spherical manifolds,
the observer dependence of the multipole expansion, and quadratic sum rules 
following from the reduction of representations. 
We show the equivalence of the Platonic cubic manifold $N2$ and the lens manifold $L(8,3)$
and investigate the statistical properties of CMB anisotropies.
The geometry of the manifold can be expressed by the Voronoi domain (see below)
which has the observer of the CMB radiation in its centre.
It turns out that the Platonic cubic geometry of the manifold $N2$
is equivalent to a special observer position in the manifold $L(8,3)$.
This manifold is not homogeneous and is thus called inhomogeneous.
In order to emphasise this point,
consider two observers where the first observer position can be mapped by
a transformation $M$ onto the second one.
Assume that the first observer determines his Voronoi domain by the
group elements $g\in H$,
then the second observer gets his Voronoi domain by the group $M g M^{-1}$.
That is a similarity transformation, or a coordinate transformation.
If $M$ and $g$ commute for all $g\in H$,
then both observers see the same Voronoi domain.
In this case, one has a homogeneous manifold.
On the other hand, if $M$ and $g$ do not commute,
one obtains observer dependent Voronoi domains and
thus an inhomogeneous manifold.
The interesting point of view with respect to cosmic topology is
that the statistical properties of the CMB radiation is observer dependent
in the inhomogeneous case.
As two examples for homogeneous manifolds,
we also analyse the lens space $L(8,1)$ and the Platonic cubic manifold $N3$.
The latter is equivalent to a manifold generated by
the binary dihedral group $D_8^*$ isomorphic to the quaternion group $Q$.

\section{Specification and equivalence of spherical  manifolds
with volume $V({\mathbb S}^3)/8$.}
\label{sec:equi}

The geometry of a spherical manifold  is not determined  by a quotient
${\mathbb S}^3/H$ if its deck group $H$ is only specified by its
group relations. 
Once we have identified a group $H$ and its action on the 3-sphere
for two geometric shapes,  we only know that both shapes may serve as 
fundamental domains under $H$ acting on the cover. 
A fundamental domain for $H$ is a subset of points on the cover such that no element of 
$g \in H,\: g\neq e$ can operate  inside the domain, 
but any point of the cover outside the domain can be reached 
by the action of $H$ on a point inside the domain.
The fundamental domain ${\cal F}$ with respect to the position $x_o$
of the observer is defined to be the set of points $x$
which satisfy
\begin{equation}
\label{Eq:Def_Voronoi}
d(x_o,x) \; \leq \; d(x_o,g(x)) \; \; \forall \; \; g \in H
\hspace{10pt} ,
\end{equation}
where $d(x,x')$ is the distance between the points $x$ and $x'$.
A fundamental domain constructed in this natural way  is called
Voronoi domain.
For historical reasons, there are several other names for such a domain
in use, for example the Dirichlet cell,
see \cite{Okabe_et_al_2000},
but we use Voronoi in the following.

How can we find out if two spherical manifolds are equivalent under homotopy?
The example of two cubic spherical manifolds shows that, even for equal 
geometric shape of the fundamental domain, their topologies, 
encoded in their homotopic boundary conditions, can be inequivalent,
i.\,e.\ they can possess different deck groups.
Homotopic equivalence requires to find a one-to-one map
between the two geometric shapes
which reproduces the homotopic boundary conditions.
We demonstrate homotopic equivalence on the Platonic cubic manifold $N2$
from \cite{KR10} versus the lens spherical manifold $L(8,3)$,
see \cite{SE34} p.\,210. 

Let us now turn to the specification of spherical manifolds
with volume $V({\mathbb S}^3)/8$,
i.\,e.\ manifolds generated by groups having 8 elements.
There are two lens spaces $L(p,q)$ with order 8,
since $p$ and $q$ have not only to be relatively prime with $1\leq q<p$.
The spaces $L(p,q)$ and $L(p',q')$ are homeomorphic
if and only if $p=p'$ and either $q=\pm q' (\hbox{mod } p)$ or
$q\,q' = \pm 1 (\hbox{mod } p)$ \cite{Gausmann_et_al_2001}.
For example, the lens spaces $L(p,q)$ and $L(p,p-q)$ are mirror images.
These restrictions leave as representations of the cyclic group $C_8$
only the lens spaces $L(8,1)$ and $L(8,3)$,
where the former is a homogeneous and the latter an inhomogeneous manifold.
A further manifold generated by a group of order 8
corresponds to the binary dihedral group $D_8^*$,
isomorphic \cite{KR09} to the quaternion group $Q$,
and admits the cubic Platonic manifold $N3$.
The lens manifold $L(8,3)$ is equivalent to the
Platonic cubic manifold $N2$.
This completes the list of manifolds with volume $V({\mathbb S}^3)/8$.

\subsection{The cubic spherical manifolds $N2$ and $N3$.}
\label{subsec:cubic}

\begin{figure}
\begin{center}
\includegraphics[width=0.7\textwidth]{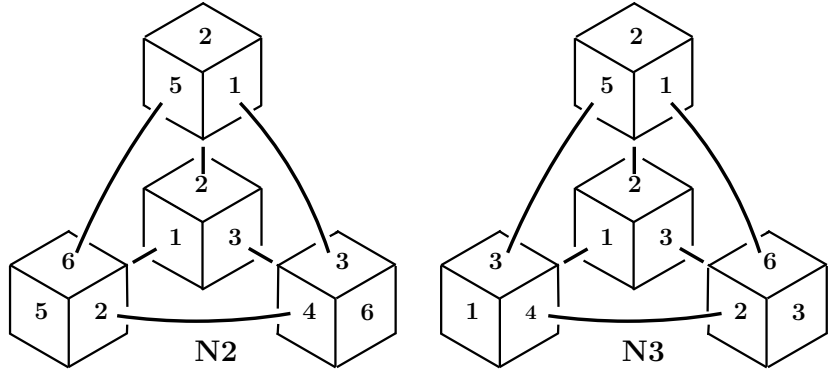} 
\end{center}
\caption{\label{fig:ncubes} The cubic manifolds $N2$ and $N3$. 
The  cubic prototile and three neighbour tiles sharing 
its faces $F1,F2,F3$. 
The four cubes are replaced by their  Euclidean counterparts  
and separated from one another. 
Visible faces are denoted by the numbers in eq.\,(\ref{e2}). 
The actions transforming the prototile into its three neighbours generate 
the deck transformations and the 8-cell tiling of ${\mathbb S}^3$. 
In the tiling, homotopic face gluing takes the form of  shared
pairs of faces $N2: F3\cup F1, F4\cup F2, F6 \cup F5$ 
and $N3: F1\cup F6, F2\cup F4, F3 \cup F5$. 
It is  marked by heavy lines or arcs.}
\end{figure}

There are two inequivalent Platonic cubic spherical manifolds $N2$ and $N3$ 
\cite{KR10}, with homotopy groups derived in \cite{EV04}. 
Their gluing is shown in figure\,\ref{fig:ncubes}. 
First we consider the gluing of the spherical cube which leads to 
the manifold $N2$.

\subsubsection*{Face gluings N2:}
\label{susubsec:face_gluing_N2}

After correction of  an  error in \cite{KR09} eq.\,(9), we have from \cite{KR10}
\begin{equation}
\label{e2}
N2:\: F3\cup F1,\; F4\cup F2,\; F6\cup F5.
\end{equation} 

\subsubsection*{Edge gluing scheme N2:}
\label{susubsec:edge_gluing_N2}
Directed edges in a single line in eq.\,(\ref{e3}) are glued. 
A bar over an edge number means that the direction of the edge is reversed
before the gluing.
\begin{equation}
\label{e3}
N2:\:\left[
\begin{array}{lll}
 1&3&4\\
2&6&\overline{9}\\
5&7&\overline{10}\\
8&11&\overline{12}\\
\end{array} \right]
\end{equation}

\begin{figure}
\begin{center}
\includegraphics[width=0.5\textwidth]{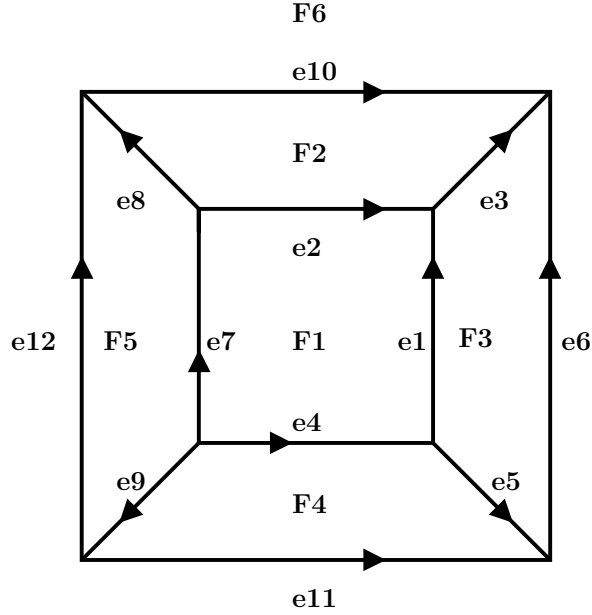} 
\end{center}
\caption{\label{fig:cubus2} 
A sketch to explain the gluing of the faces ($Fi$) and the edges ($ei$) 
of the spherical cubic manifolds $N2$ and $N3$.}
\end{figure}

The elements of the rotation group
\begin{equation}
\label{Ref:su(4)}
\hbox{SO}(4,\mathbb{R}) \, \sim \,
(\hbox{SU}^l(2,\mathbb{C})\times \hbox{SU}^r(2,\mathbb{C}))/\{\pm (e,e)\}
\end{equation}
are denoted as pairs $(g_l, g_r)$.
They act on the points $u \in \hbox{SU}(2,\mathbb{C})$ of the 3-sphere
$\mathbb{S}^3 \equiv \hbox{SU}(2,\mathbb{C})$ by
\begin{equation}
\label{Ref:act}
(g_l, g_r): u \rightarrow g_l^{-1}u g_r.
\end{equation}

\subsubsection*{Group $H={\rm deck}(N2)$:}
\label{susubsec:group_N2}
Another description of this manifold is given by the identification of
the points on the covering space ${\mathbb S}^3$ by the group 
$H={\rm deck}(N2)$ which is the cyclic group $C_8$ generated by the element
(an irrelevant minus sign in \cite{KR10} has been dropped)
\begin{equation}
\label{Eq:generator_group_N2}
g \,:=\, \left(g_l,g_r\right) \;=\; 
\left(\begin{array}{cc}
\left[
\begin{array}{ll}
\overline{a}&0\\
0&a
\end{array}
\right],
&
\left[\begin{array}{ll}
0&a^3\\
a&0
\end{array}
\right] 
\end{array}
\right)
\hspace{10pt}
\end{equation}
with $a=\exp(2\pi \hbox{i}/8)$,
and $g_l,g_r \in \hbox{SU}(2,{\mathbb C})$.
Then the corresponding manifold is invariant under
\begin{equation}
\label{Eq:action_group_N2}
u \rightarrow u_n \; = \; \left(g_l^{-1}\right)^n\,u\,\left(g_r\right)^n \;,\;\; n=1,..,8
\hspace{10pt}
\end{equation}
of all points
\begin{equation}
\label{Eq:coordinates_u}
u \; = \;
\left[
\begin{array}{ll}
z_1&z_2\\
-\overline{z}_2&\overline{z}_1
\end{array}
\right]
\; = \;
\left[
\begin{array}{rr}
x_0-\hbox{i}x_3& -x_2-\hbox{i}x_1\\
x_2-\hbox{i}x_1&x_0+\hbox{i}x_3
\end{array}
\right]
\in \hbox{SU}(2,{\mathbb C}) \equiv {\mathbb S}^3
\hspace{10pt}.
\end{equation}
Transcribing the same action of this group $C_8$ into real notation we get 
\begin{equation}
\label{Eq:action_group_N2_real}
\vec{x} \; = \; (x_0,x_1,x_2,x_3)^T\in {\mathbb S}^3 
\rightarrow \vec{x}_n  \; = \; \left(R_g\right)^n\,\vec{x}\;,\;\; n=1,..,8
\hspace{10pt}
\end{equation}
with the generator of the group
\begin{equation}
\label{Eq:generator_N2_real}
R_g \; = \;
\left[
\begin{array}{rrrr}
0&1&0&0\\
0&0&0&-1\\
1&0&0&0\\
0&0&1&0
\end{array}
\right]
\in \hbox{SO}(4,{\mathbb R})
\hspace{10pt},
\end{equation}
see also table~3 in \cite{KR09}.

Now we carry out the analogous considerations for the manifold $N3$.
\subsubsection*{Face gluings N3:}
\label{susubsec:face_gluing_N3}

Opposite faces of the cube are glued, 
\begin{equation}
\label{Eq:face_gluing_N3}
N3:\: F1\cup F6,\; F2\cup F4,\; F3\cup F5.
\end{equation}

 \subsubsection*{Edge gluing scheme N3:}
\label{susubsec:edge_gluing_N3}
Directed edges in a single line are glued.
\begin{equation}
\label{Eq:edge_gluing_N3}
N3:\:\left[
\begin{array}{lll}
 1&8&11\\
2&\overline{6}&\overline{9}\\
3&4&\overline{12}\\
5&\overline{7}&\overline{10}\\
\end{array} \right]
\end{equation}

\subsubsection*{Group $H={\rm deck}(N3)$:}
\label{susubsec:group_N3}

The group $H={\rm deck}(N3)$ is the binary dihedral group $D^*_8$
generated by the two elements 
\begin{eqnarray}
\label{Eq:generators_group_N3}
\nonumber
& g_1 \,:=\, \left(g_{l1},e\right) 
\; , \;\; 
g_2 \,:=\, \left(g_{l2},e\right) \\ 
&\\
& \nonumber \hbox{with}\;
g_{l1}\;=\;
\left[
\begin{array}{ll}
0&\hbox{i}\\
\hbox{i}&0
\end{array}
\right],
\;
g_{l2}\;=\;
\left[
\begin{array}{ll}
\hbox{i}&0\\
0&-\hbox{i}
\end{array}
\right],
\;
e\;=\;
\left[
\begin{array}{ll}
1&0\\
0&1
\end{array}
\right]
\hspace{10pt},
\end{eqnarray}
or equivalently  by
\begin{equation}
\label{Eq:generator_N3_real}
R_{g_1}  \, = \, 
\left[
\begin{array}{rrrr}
0&-1&0&0\\
1&0&0&0\\
0&0&0&-1\\
0&0&1&0
\end{array}
\right]
\,, \,\,
R_{g_2}  \, = \,
\left[
\begin{array}{rrrr}
0&0&0&-1\\
0&0&-1&0\\
0&1&0&0\\
1&0&0&0
\end{array}
\right]
\in \hbox{SO}(4,{\mathbb R})
\hspace{5pt}.
\end{equation}

\subsection{Transformation of the observer position}

Let us now address the question how the group $g\in H$ transforms
under a change of the observer position,
whereby each observer naturally puts his position at the origin of his
coordinate system.
The behaviour under such transformations will determine
whether a spherical manifold is homogeneous or inhomogeneous.
By applying an arbitrary transformation $q$ to the coordinates
\begin{equation} 
\label{Eq:trafo_coordinate}
u \rightarrow u'= u\, q \; \; , \; \; q \in \hbox{SU}(2,{ \mathbb C})
\hspace{10pt} ,
\end{equation}
we can transform the origin of the coordinate system to every point on
${\mathbb S}^3$ with the isometry $q$.
By the transformation $q$, an observer sitting at $u=q^{-1}$
is shifted to the centre of the new coordinate system $u'=e$.
Now consider a given point whose coordinates with respect to
two observers $o$ and $o'$ are related by $u' = u q$.
The group elements of the deck transformation with respect to the observer $o$ 
is given by $g_i=(g_{li},g_{ri})$, $i=1, 2, \dots$.
Using these deck transformations we get for every point $u$ on the 3-sphere 
points $\tilde{u}_i$ that are to be identified, 
i.\,e.\ $\tilde{u}_i\equiv (g_{li})^{-1}\,u\,g_{ri}$.
Transforming these points into the observer system $o'$
we get
\begin{eqnarray} \nonumber
\tilde{u}_i \rightarrow \tilde{u}'_i = \tilde{u}_i\,q 
& = &(g_{li})^{-1}\,u\,g_{ri}\,q \\
\label{Eq:trafo_coordinate_general_group}
& = & (g_{li})^{-1}\,u\,q\,(q^{-1}\,g_{ri}\,q)
= (g_{li})^{-1}\,u'\,(q^{-1}\,g_{ri}\,q)
\hspace{10pt}.
\end{eqnarray}
The observer $o'$ uses the equation $\tilde{u}'_i= (g'_{li})^{-1}\,u'\,g'_{ri}$
to identify points on the 3-sphere. 
Comparing this equation with eq.\,(\ref{Eq:trafo_coordinate_general_group}),
one gets the deck transformations
\begin{equation}
\label{Eq:trafo_group}
g'_i \; = \; (g'_{li},g'_{ri}) \; = \;(g_{li},q^{-1}\,g_{ri}\,q)
\hspace{10pt} , \hspace{10pt}
i=1, 2,\dots
\hspace{10pt} ,
\end{equation}
with respect to the observer $o'$.
Since the coordinate transformation $q$ is given by right action,
the left action $g_{li}$ of a deck transformation $g_i$ does not change,
but the right action $g_{ri}$ of a deck transformation $g_i$ 
in general changes under a coordinate transformation $q$.

In case of the manifold $N3$,
the group elements $g_i=(g_{li},e)=g'_i$, $i=1,...,8$,
see eq.\,(\ref{Eq:generators_group_N3}),
do not change under the transformation (\ref{Eq:trafo_coordinate})
because of $g_{ri}=e$.
The invariance of the group elements implies
that the same fundamental domain is obtained
for every choice of the coordinate system.
Such manifolds are called homogeneous, see e.\,g.\ p.\,230 in \cite{WO84}.
In the case of the manifold $N2$
the transformation (\ref{Eq:trafo_coordinate}) changes
the corresponding group elements $g_i=(g_{li},g_{ri})\rightarrow
g'_i=(g_{li},g'_{ri})=(g_{li},q^{-1}\,g_{ri}\,q)$, $i=1,...,8$.
Because of $q^{-1}\,g_{ri}\,q \neq g_{ri}$, in general,
a different choice of the observer position usually result 
in another shape of the Voronoi domain, see eq.\,(\ref{Eq:Def_Voronoi}).
Such a manifold is called an inhomogeneous manifold.
These changes in the shape of the fundamental domain are
illustrated in fig.\,\ref{fig:fundamental_cell_N2_vs_L83},
where the position of the observer is shifted using the parameterisation
\begin{equation}
\label{Eq:coordinate_q_rho_alpha_epsilon}
q(\rho,\alpha,\epsilon) \; = \;\left[ \begin{array}{l@{\quad}l@{\quad}}
\cos(\rho)\,\exp(-\hbox{i}\alpha)
&-\hbox{i}\sin(\rho)\,\exp(-\hbox{i}\epsilon) \\
-\hbox{i}\sin(\rho)\,\exp(+\hbox{i}\epsilon)
&\cos(\rho)\,\exp(+\hbox{i}\alpha)
\end{array} \right]
\end{equation}
with $\rho \in [0,\frac{\pi}{2}]$, $\alpha, \epsilon \in [0,2\pi]$.

\subsection{Relation of the  lens manifold $L(8,3)$ and $N2$.}
\label{subsec:rela}

To discuss the relation between these two spherical manifolds
we look at their representations. 
With the generator $g=(g_l, g_r)$ of the group $H=C_8$
for the cubic manifold $N2$ given in eq.\,(\ref{Eq:generator_group_N2}), 
it is easy to transform the generator $g$ to diagonal form
$g_d:=(\delta_l, \delta_r)$,
\begin{eqnarray}
\label{Eq:trafo_generator_N2_L83}
\delta_l & := & g_l \; = \;
\left[\begin{array}{ll}
\overline{a}&0\\0&a
\end{array}\right]
\; = \;
\left[
\begin{array}{ll}
\exp(-\hbox{i}\frac{\pi}{4})&0\\
0&\exp(\hbox{i}\frac{\pi}{4})\\
\end{array}
\right] \; ,
\\ \nonumber
\delta_r & := & q^{-1}\, g_r \, q
\; = \;
q^{-1} \left[\begin{array}{ll}
0&a^3\\a&0
\end{array}\right] q
\; = \;
\left[
\begin{array}{ll}
\exp(-\hbox{i}\frac{\pi}{2})&0\\
0&\exp(\hbox{i}\frac{\pi}{2})\\
\end{array}
\right]\; \; ,
\end{eqnarray}
where the coordinate shift $q$, eq.\,(\ref{Eq:coordinate_q_rho_alpha_epsilon}),
has to be chosen as
\begin{equation}
\label{Eq:trafo_rho}
q \, = \,
q\left(\rho=\frac \pi 4, \alpha=\frac{7\pi}8, \epsilon=\frac{3\pi}8\right)
\, = \,
\frac{1}{\sqrt 2}
\left[
\begin{array}{ll}
-\exp(\hbox{i}\frac{\pi}{8})& -\exp(\hbox{i}\frac{\pi}{8})\\
\exp(-\hbox{i}\frac{\pi}{8})& -\exp(-\hbox{i}\frac{\pi}{8})\\
\end{array}
\right] \, .
\end{equation}
If we define a transformation of coordinates $u$ according to
\begin{eqnarray}
 \label{Eq:trafo_N2_L83}
u \rightarrow  u'
=\left[
\begin{array}{rr}
x'_0-\hbox{i}x'_3& -x'_2-\hbox{i}x'_1\\
x'_2-\hbox{i}x'_1&x'_0+\hbox{i}x'_3\\
\end{array}
\right]=u \, q
\hspace{10pt} ,
\end{eqnarray}
the action of the generator $g_d=(\delta_l,\delta_r)$ on the new coordinates
$u'$ follows from eq.\,(\ref{Eq:trafo_group})
\begin{equation}
\label{e6}
u' \rightarrow \delta_l^{-1} u' \delta_r
\hspace{10pt} .
\end{equation}
This action can be expressed in terms of the complex coordinates  
$(z_1', z_2')$ of $u'$ by
\begin{equation}
\label{e7}
g_d: (z_1', z_2') \rightarrow  
(z_1'\,\overline{a},\, z_2' \,  a^3)
\; = \;
\left(z_1'\exp\left(\hbox{i}\frac{-2\pi}{8}\right),\,
z_2'\exp\left(\hbox{i}\frac{3\cdot 2\pi}{8}\right)\right)
\end{equation}
with $a=\exp(2\pi \hbox{i}/8)$.
Using the parameterisation~(\ref{Eq:coordinate_q_rho_alpha_epsilon}) 
also for the coordinate $u$,
the action of the generator $g_d=(\delta_l, \delta_r)$ is given by
\begin{equation}
\label{Eq:action_generator_N2_L83}
g_d: u(\rho, \alpha, \epsilon)  \rightarrow 
u\left(\rho, \alpha + \frac{2 \pi}{8}, \epsilon- \frac{3 \cdot 2 \pi}{8}\right)
 \; \; .
\end{equation}
The complex representation (\ref{e6}) of the cubic generator $g_d$
of the manifold $N2$ corresponds in the (real) classification by
representations of Wolf \cite{WO84} p.\,224 exactly to the
spherical lens space $L(n,k)=L(8,3)$.
From this algebraic equivalence of the representations of $C_8$ for the
lens manifold $L(8,3)$ according to \cite{WO84} and
for the cubic manifold $N2$, eq.\,(\ref{e6}),
we conclude that there must exist a one-to-one geometric map of their
fundamental domains.
Their homotopy and deck group $H$ must coincide.

\begin{figure}
\vspace*{0pt}
\begin{minipage}{17.0cm}
\begin{minipage}{9.5cm}
{
\includegraphics[width=8.5cm]{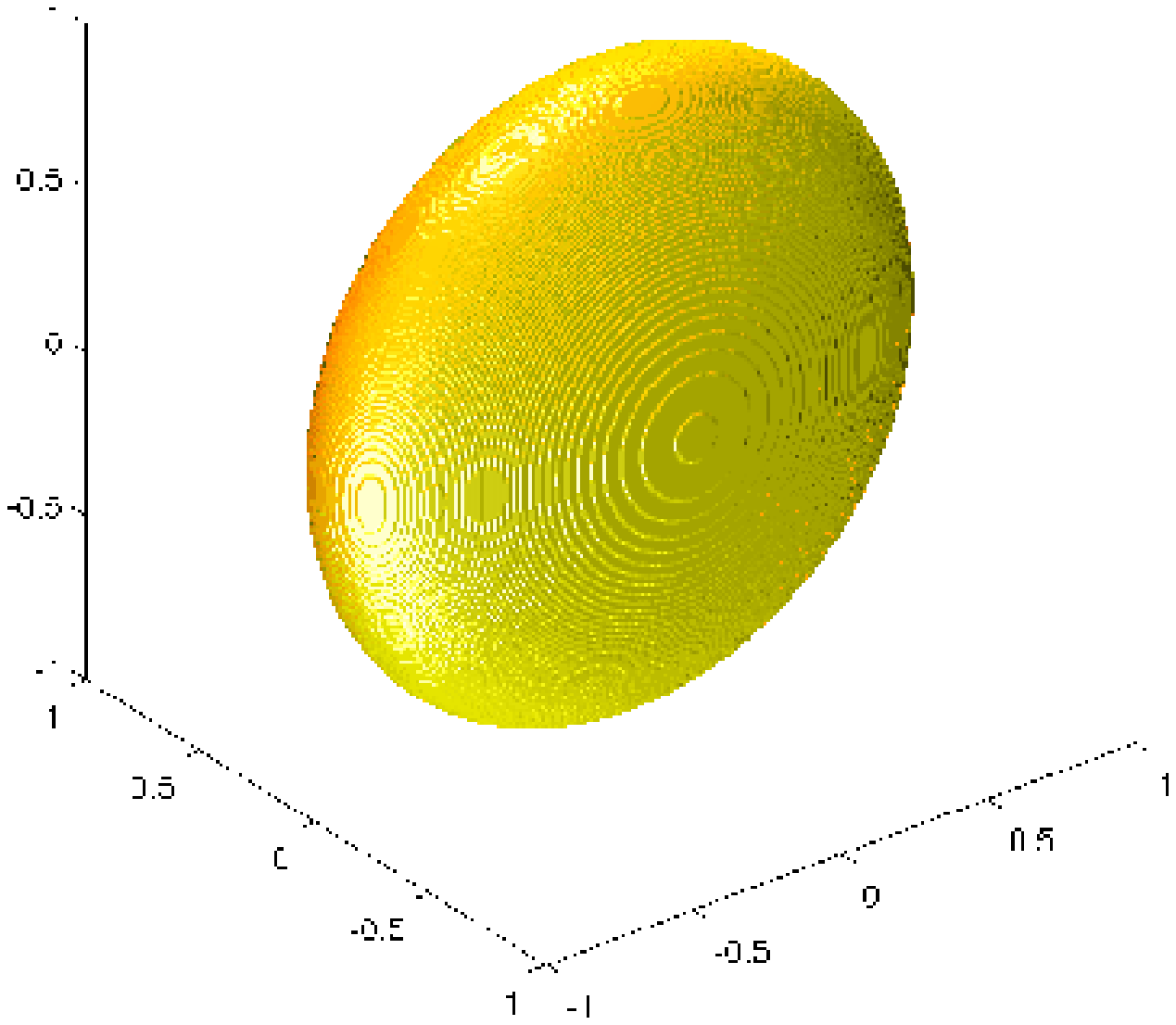}
\put(-200,165){a) $\rho=0.0$}
\vspace*{0pt}
}
{
\includegraphics[width=8.5cm]{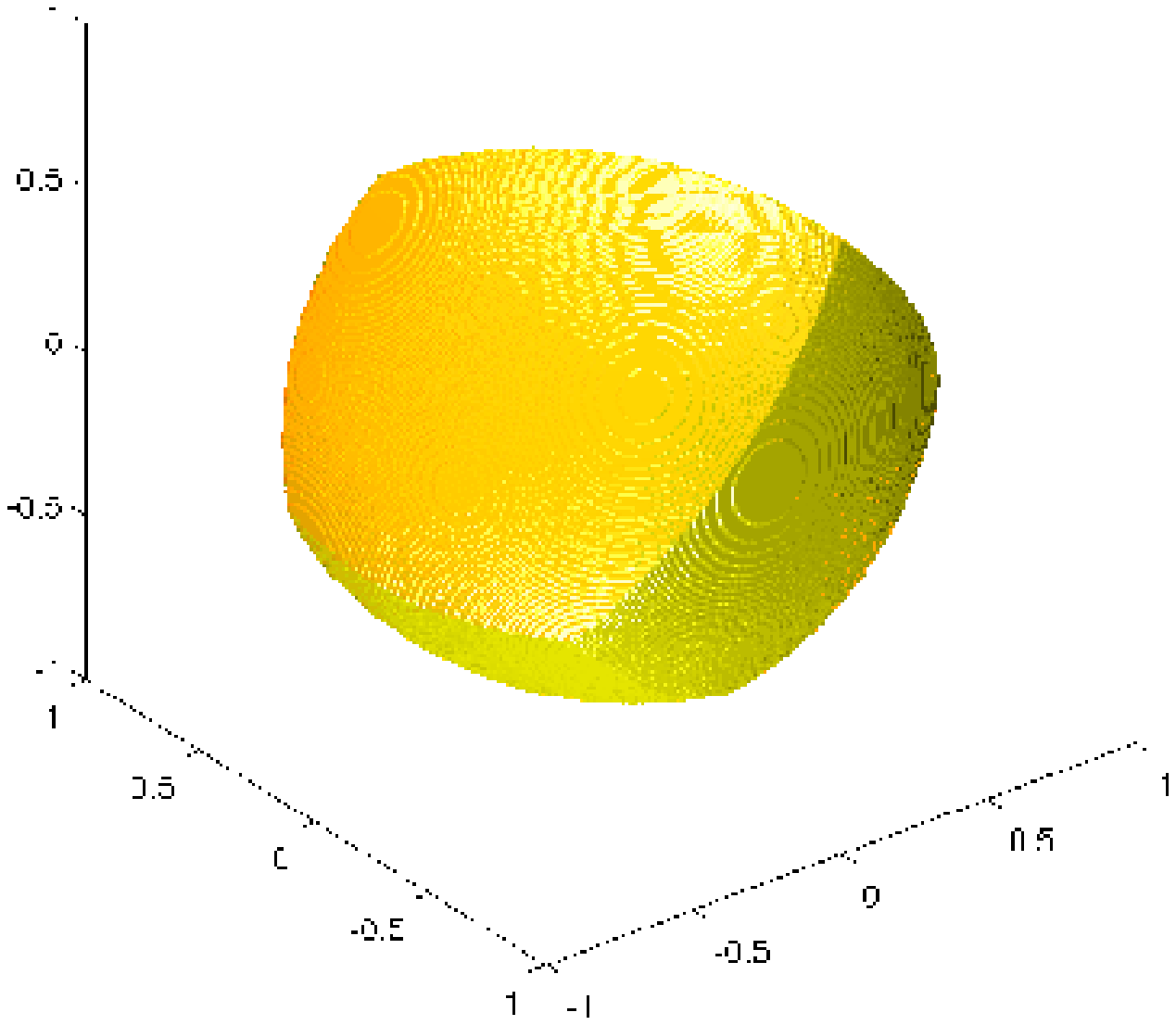}
\put(-200,165){c) $\rho=0.15\, \pi$}
\vspace*{0pt}
}
\end{minipage}
\hspace*{-55pt}
\begin{minipage}{9.5cm}
{
\includegraphics[width=8.5cm]{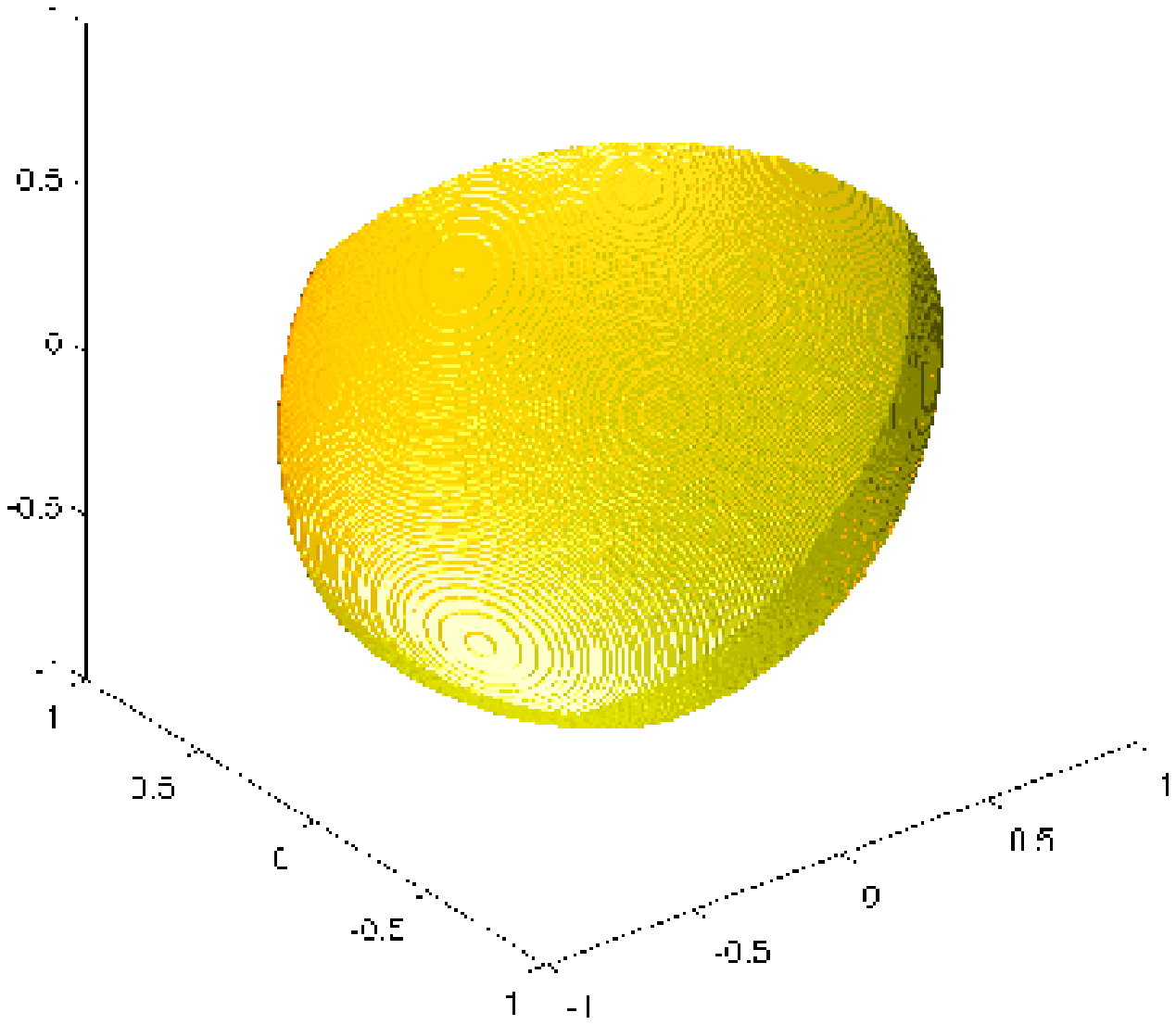}
\put(-200,165){b) $\rho=0.08\, \pi$}
\vspace*{0pt}
}
{
\includegraphics[width=8.5cm]{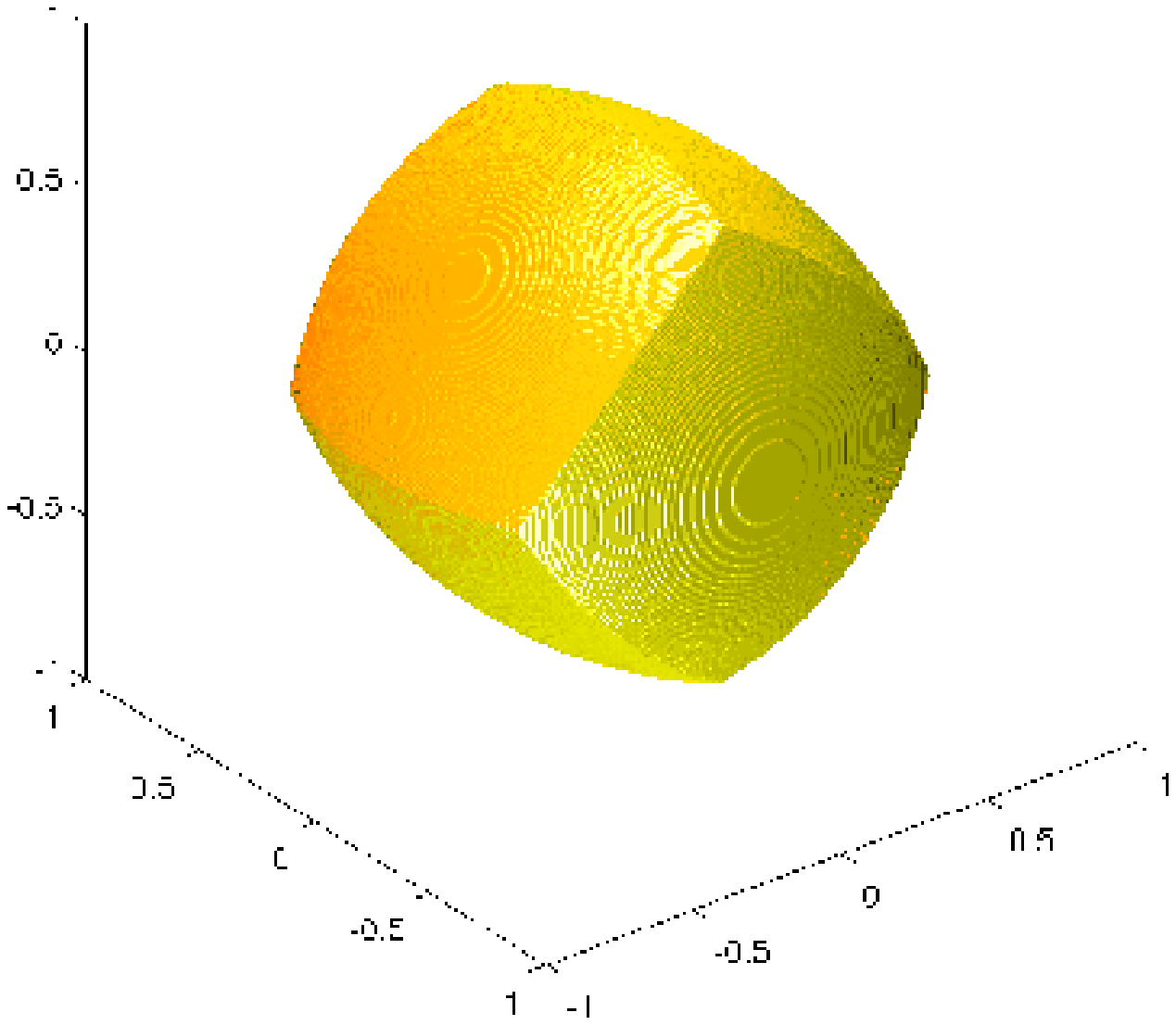}
\put(-200,165){d)  $\rho=0.25\, \pi$}
\vspace*{0pt}
}
\end{minipage}
\end{minipage}
\caption{\label{fig:fundamental_cell_N2_vs_L83}
The Voronoi domains of the manifold $L(8,3) \equiv N2$ 
at four different positions of the observer $q^{-1}$ are shown
using the projection to ${\mathbb R}^3$.
Within the parameterisation of the observer 
point, see eq.\,(\ref{Eq:coordinate_q_rho_alpha_epsilon}),
the coordinate $\rho$ is varied. 
In doing so $\alpha, \epsilon = 0$ have been chosen.
}
\end{figure}

In the following, we choose for the generator of the cyclic group $C_8$
the equivalent diagonalised generator
$g=(\tilde\delta_l, \tilde\delta_r) :=
(\overline{\delta}_r,\overline{\delta}_l)$
that is
\begin{eqnarray}
\label{Eq:generator_N2_L83_v2}
&&\tilde\delta_l =
\left[
\begin{array}{ll}
\exp(\hbox{i}\frac{\pi}{2})&0\\
0&\exp(-\hbox{i}\frac{\pi}{2})\\
\end{array}
\right]
\hspace{10pt} , \hspace{10pt}
\tilde\delta_r=
\left[
\begin{array}{ll}
\exp(\hbox{i}\frac{\pi}{4})&0\\
0&\exp(-\hbox{i}\frac{\pi}{4})\\
\end{array}
\right]\; \; \; .
\end{eqnarray}
This generator and the generator $g_d:=(\delta_l, \delta_r)$
describe isospectral manifolds.
In \cite{Ikeda_1980} the following Theorem is proven:
``If two 3-dimensional spherical space forms are isospectral,
then they are isometric.''
Thus, the two generators lead to equivalent manifolds.

The action of the generator $g=(\tilde\delta_l, \tilde\delta_r)$,
eq.\,(\ref{Eq:generator_N2_L83_v2}),
changes the negative sense of the rotation in the $x_1$-$x_2$-plane
described in eq.(\ref{Eq:action_generator_N2_L83}) into a positive sense
\begin{equation}
\label{Eq:action_generator_N2_L83_v2}
g: u(\rho, \alpha, \epsilon)  \rightarrow 
u\left(\rho, \alpha + \frac{2 \pi}{8}, \epsilon+ \frac{3 \cdot 2 \pi}{8}\right)
 \; \; .
\end{equation}
The Voronoi domain of the group generated by
(\ref{Eq:action_generator_N2_L83_v2}) is
pictured in fig.\,\ref{fig:fundamental_cell_N2_vs_L83}a.
The Voronoi domain is shown as a projection onto ${\mathbb R}^3$.
This projection is defined as simply omitting the component $x_0$
in the vector $\vec x$, see eq.\,(\ref{Eq:action_group_N2_real}).
The Voronoi cell is computed with respect to an observer at the
origin $\vec x_0=(1,0,0,0)$ by transforming the group elements
of $L(8,3)$ using eq.\,(\ref{Eq:trafo_group}).
Now choosing in eq.\,(\ref{Eq:coordinate_q_rho_alpha_epsilon})
for the parameters $\alpha, \epsilon=0$,
we obtain for $\rho=0.08 \pi,\; 0.15 \pi$, 
and $0.25\pi$ new generators $g'=(\tilde\delta_l, q^{-1} \tilde\delta_r q)$.
The outcome of these are the Voronoi domains
shown in  figures \ref{fig:fundamental_cell_N2_vs_L83}b-d.
For three values of $\rho$ the actions of the deck transformations 
result in fundamental cells with very special shapes.
For $\rho=0$ the Voronoi domain is given by 
a spherical lens and for $\rho=0.25 \pi$ by the spherical Platonic cube $N2$,
as revealed by eq.\,(\ref{Eq:trafo_rho}) using
$\sin\frac\pi 4=\cos\frac\pi 4=\frac 1{\sqrt 2}$.
In the case of $\rho=0.5 \pi$ the generator of the group transforms to
$g'=(\tilde\delta_l, \overline{\tilde\delta}_r) =
(\overline{\delta}_r,\delta_l)$
again resulting in the geometric shape of a spherical lens,
but the action of this generator on 
the coordinate $u'$ have been exchanged compared to 
eq.~(\ref{Eq:action_generator_N2_L83_v2}),
$g':u'(\rho, \alpha, \epsilon)\rightarrow
u'(\rho, \alpha+ \frac{3 \cdot 2 \pi}{8}, \epsilon+ \frac{2 \pi}{8})$.

\subsection{The lens manifold $L(8,1)$.}
\label{subsec:group_C8_L81}

In addition to $L(8,3) \equiv N2$ there is another specification of the 
abstract group $C_8$ which results in the lens space $L(8,1)$.
The generator of this group can be chosen as
\begin{equation}
\label{Eq:generator_group_L81}
g \, := \, \left(g_l,e\right)
\hspace{10pt} \hbox{with} \hspace{10pt}
g_l \; := \; 
\left[\begin{array}{ll}
\exp\left(\hbox{i}\frac{\pi}{4}\right)&0\\
0&\exp\left(-\hbox{i}\frac{\pi}{4}\right)
\end{array}\right]
\hspace{10pt} .
\end{equation}
The action of the generator (\ref{Eq:generator_group_L81})
on the coordinate $u$ is given by
\begin{equation}
\label{Eq:action_generator_L81}
g: u(\rho, \alpha, \epsilon)  \rightarrow 
u\left(\rho, \alpha + \frac{2\pi}{8}, \epsilon + \frac{2\pi}{8} \right)
\hspace{10pt} .
\end{equation}
This realisation of the cyclic group $C_8$ describes a homogeneous manifold.
The Voronoi domain of this manifold is a spherical lens 
similar to the case $\rho=0.0$ of the manifold $L(8,3)$.
But the actions of these two cyclic groups $C_8$ are not equivalent
as the comparison of (\ref{Eq:action_generator_L81}) with
(\ref{Eq:action_generator_N2_L83_v2}) shows.

\section{Eigenmodes of spherical manifolds and sums related to them.}
\label{sec:observ}

\subsection{Harmonic analysis on a spherical manifold.}
\label{subsec:harmon}

From deck groups acting on the 3-sphere $\mathbb{S}^3$
one can go globally to harmonic analysis on the 3-sphere. 
One can establish functional analysis globally on the cover by determining 
a complete set of functions on the 3-sphere, each of them invariant under 
the action of the chosen group $H={\rm deck}({\cal M})$. 
The same set of functions, restricted to the prototile ${\cal M}$, locally
obeys its homotopic boundary conditions and so yields a local basis on it. 

Since the group $H$ is a  subgroup
of $\hbox{SO}(4,\mathbb{R})$, it also commutes with the Euclidean Laplacian $\Delta$
on $\mathbb{E}^4$. 
By restricting the set of functions to those that vanish under $\Delta$,
termed harmonic, we are doing harmonic analysis
on the manifold ${\cal M}$.

For all spherical manifolds, the global covering 3-sphere $\mathbb{S}^3$ and
its eigenmodes form an arena of comparison.
By extending any $H$-invariant polynomial to a polynomial function
on the 3-sphere,
we circumvent the need to explicitly identify any point or functional value
on the sphere with a point or functional value on the chosen single
topological manifold ${\cal M}$.
The Wigner polynomials $D^j_{m_1, m_2}(u),\: u \in \hbox{SU}(2,\mathbb{C})$,
introduced by Wigner \cite{WI59} for the analysis on $\hbox{SU}(2,\mathbb{C})$,
for all degrees $2j=0, 1,2,..., \infty$ form an orthogonal and complete set of
harmonic functions on the 3-sphere $\mathbb{S}^3$ \cite{KR10}.
The general action of
$\hbox{SO}(4,\mathbb{R}) \sim
(\hbox{SU}^l(2,\mathbb{C}) \times \hbox{SU}^r(2,\mathbb{C})) / \{\pm (e, e)\}$
with elements $g= (g_l,g_r)$ on Wigner polynomials and its representation
from eqs.\,(\ref{Ref:su(4)}) and (\ref{Ref:act}) are
\begin{eqnarray}
 \label{x1}
\nonumber
(T_{(g_l,g_r)} D^j_{m_1m_2})(u) 
&:=&D^j_{m_1m_2}(g_l^{-1}ug_r)\\
&=&\sum_{m_1'm_2'}D^j_{m_1'm_2'}(u) \, D^j_{m_1m_1'}(g_l^{-1}) \, D^j_{m_2'm_2}(g_r).
\end{eqnarray}
Therefore, we can write the basis of the harmonic analysis on ${\cal M}$ to 
its $H$-invariant subbasis by projection from Wigner polynomials.  

We introduce the Wigner basis, compare \cite{KR10},
with $\beta=2j+1$,
\begin{equation}
\label{Eq:wigner_basis}
\psi(j,m_1,m_2)(u) \, = \, \frac{2j+1}{\sqrt{2\,\pi^2}} \;
\langle j-m_1jm_1|0 0\rangle \; D^j_{-m_1m_2}(u) 
\hspace{10pt} ,
\end{equation}
which is normalised on the 3-sphere $\mathbb{S}^3$.
In the following, we use the spherical basis
\begin{equation}
\label{x2}
\psi(j,l,m)(u) \; = \; \sum_{m_1} \psi(j,m_1,m_2)(u) \;
\langle jm_1jm_2|lm\rangle
\end{equation}
which is related to the Wigner basis also by
\begin{equation}
\psi(j,m_1,m_2)(u)\; = \;
\sum_{l}\psi(j ,l, m)(u) \; \langle jm_1jm_2|lm\rangle
\hspace{10pt} ,
\end{equation}
where $\langle jm_1jm_2|lm\rangle$ are the Clebsch-Gordan coefficients 
which obey $0\leq l \leq 2j$ and $m = m_1+m_2$.
More details on these coefficients are given e.\,g. in \cite{ED64}.
The spherical basis eq.\,(\ref{x2}) has the property
that under the conjugation action of $\hbox{SU}^c(2,\mathbb{C})$,
it transforms as
\begin{eqnarray}
\label{x3}
\nonumber
h=(g,g)\: :
(T_{g,g}\psi)(j,l,m )(u)&=&\psi(j,l,m)(g^{-1}ug) \\
&=& \sum_{m'} \psi(j,l,m')(u) \; D^l_{m' m}(R(g)) \; \; , 
\end{eqnarray}
where $R(g)$ is the rotation w.r.t.\ the coordinates $(x_1,x_2,x_3)$
that corresponds to $g \in \hbox{SU}^C(2,\mathbb{C})$. 
With a set $(\chi, \theta, \phi)$ of coordinates for $u$
\cite{AU05b,KR10}, the spherical basis is proportional to the
standard spherical harmonics $Y^l_m(\theta, \phi)$
and is given by $\psi(j,l,m)(\chi, \theta, \phi)= \hbox{i}^l 
R^{l}_{\beta}(\chi)Y^l_m(\theta, \phi) $ with
$$
R^{l}_{\beta}(\chi)=(-\hbox{i})^l \sqrt{\frac{2\beta^2}{(2l+1)}}
\sum_{m_1}\langle j-m_1jm_1|0 0\rangle\langle j-m_1jm_1|l 0\rangle D^j_{m_1 m_1}(\chi,0,\chi)
\hspace{10pt},
$$
see \cite{Gundermann_2005}.
A different phase convention for the radial function is used
as in \cite{Gundermann_2005}.
This spherical basis is very convenient for the multipole expansion 
of invariant polynomials.

What happens to the CMB multipole amplitudes  when we adopt on the manifold  
a general position of the observer?

We use Cartesian coordinates $x$ in Euclidean 4-space and $2 \times 2$  
matrix coordinates $u(x)$ and write down the general transformation 
$u \rightarrow u'=u\: q,\: x \rightarrow x'=x\:R(q)$
of the coordinates from a special point (for example the centre of a
spherical space) to a general point.

Then we can pass to polynomial functions (Wigner or spherical) and give on the 
3-sphere their  transformation law as a function of $(q,\: R(q))$. 
We focus on the spherical basis and show how the  multipole expansion in a 
given topological model is transformed under change of the observer position 
into another multipole expansion.

Our initial basis on the 3-sphere are linear combinations of Wigner polynomials
in initial coordinates $u$, convenient for the initial observer or
suggested by a simple decomposition under $H$.
The most general shift of the observer to new coordinates $u'$ is
given by eq.\,(\ref{Eq:trafo_coordinate}).
The unitary matrix transformation $u \rightarrow u'= u\; q$ yields a rotation 
$R(q): x'=x\: R(q)$, applied to the row of coordinates 
$x=(x_0,x_1,x_2,x_3)$ of Euclidean space $\mathbb{E}^4$, given by 
\begin{eqnarray}
\label{gh2b} 
u=\left[
\begin{array}{rr}
 z_1& z_2\\
-\overline{z}_2& \overline{z}_1\\
\end{array} 
\right]  
=\left[
\begin{array}{rr}
 x_0-\hbox{i}x_3& -x_2-\hbox{i}x_1\\
x_2-\hbox{i}x_1&x_0+\hbox{i}x_3\\
\end{array} 
\right],\:  
 q=\left[
\begin{array}{rr}
 a& b\\
-\overline{b}& \overline{a}\\
\end{array} 
\right],  
\\ \nonumber 
R(q)= \frac{1}{2}
 \left[
\begin{array}{rrrr}
(a+\overline{a})&\hbox{i}(b-\overline{b})&-(b+\overline{b})&\hbox{i}(a-\overline{a})\\
-\hbox{i}(b-\overline{b})&-(a+\overline{a})& \hbox{i}(a-\overline{a})&-(b+\overline{b})\\
(b+\overline{b})&-\hbox{i}(a-\overline{a})&-(a+\overline{a})&-\hbox{i}(b-\overline{b})\\
-\hbox{i}(a-\overline{a})&(b+\overline{b})&\hbox{i}(b-\overline{b})& (a+\overline{a})\\
\end{array}
\right].
\end{eqnarray}

\subsection{Observer- and multipole dependence in the spherical basis.}
\label{subsec:observ}

Initially, the coordinates on the 3-sphere refer to the point $(1,0,0,0)$,
$u=e$.
We now pass with eq.\,(\ref{gh2b}) to a general position of the observer.

Our approach to the dependence of the multipole analysis on
the observer position goes as follows:
We break the general right transformation $(e,q)$ of the coordinates
into a diagonal part $\lambda$ and two rotations from $\hbox{SO}(3,\mathbb{R})$.
We use the bracket notation. 
The matrix $q\in \hbox{SU}(2,\mathbb{C})$ has a diagonal decomposition
\begin{equation}
\label{gh9}
q = r \lambda r^{-1} \; \; .
\end{equation}
Now for elements $(g_l, g_r) \in \hbox{SO}(4,\mathbb{R}) \sim
(\hbox{SU}^l(2,\mathbb{C}) \times \hbox{SU}^r(2,\mathbb{C}))/(\pm(e,e))$,
we have the identity
\begin{equation}
\label{gh10}
(r,r)(e, \lambda)(r^{-1},r^{-1})= (e, r \lambda r^{-1})= (e,q) \; \; .
\end{equation}
We consider the three factors on the left hand side and compute their
matrix elements in the representation of $\hbox{SO}(4,\mathbb{R})$. 

(i) In the spherical basis $\psi(j,l,m)(u), \beta=2j+1$, we have 
\begin{equation}
\label{gh11}
\langle j l m|T_{(r,r)}|jl'm'\rangle = \delta_{l l'} D^l_{mm'}(R(r)),
\end{equation}
where $R(r)$ is the rotation from $\hbox{SO}(3,\mathbb{R})$ acting on
$(x_1,x_2,x_3)$ that corresponds to $r \in \hbox{SU}(2,\mathbb{C})$.

(ii) In the Wigner polynomial basis $D^j_{m_1,m_2}(u)$
we have for the diagonal part
\begin{eqnarray}
\label{gh12}
&&\lambda = 
\left[\begin{array}{ll}
\exp(\hbox{i}\alpha/2)&0\\
0&\exp(-\hbox{i}\alpha/2)
\end{array}\right] \; , \;
\cos(\alpha/2)={\rm Trace}(q)/2\;,
\\ \nonumber 
&& \langle jm_1m_2|T_{(e,\lambda)}|jm_1'm_2'\rangle =
\delta_{m_1m_1'}\delta_{m_2m_2'} \exp(\hbox{i}\alpha (m_1-m_2)) \; \; .
\end{eqnarray}
The diagonal part $\lambda$ is completely determined by the trace.
Transforming these matrix elements into the spherical basis eq.\,(\ref{x2})
gives with eq.\,(\ref{gh12})
\begin{eqnarray}
\label{gh13}
\nonumber 
&&\langle j l m|T_{(e,\lambda)}|jl'm'\rangle \\
&& \hspace{10pt} = \sum_{m_1m_2} \langle lm|j-m_1jm_2\rangle \exp(\hbox{i}\alpha (m_1-m_2))
\langle j-m_1jm_2|l'm'\rangle
\hspace{10pt}. 
\end{eqnarray}
where two sign factors cancel.
Now we combine the three factors from eq.\,(\ref{gh10}) to obtain the
overall matrix elements in the spherical basis as
\begin{eqnarray}
\label{gh14}
\nonumber 
&&\langle jlm|T_{(e,q)}|jl'm'\rangle \\
&& \hspace{10pt} = \sum_{m''} D^l_{mm'}(R(r)) 
\langle j l m'|T_{(e,\lambda)}|jl'm''\rangle
D^{l'}_{m''m'}(R^{-1}(r))
\end{eqnarray}
with the middle matrix elements given in eq.\,(\ref{gh13}).
The matrix elements in the middle are the essential part of the transformation 
to a new observer position.
Under the rotations $R(r), R(r^{-1})$, the bases in multipole
form have standard properties.

{\bf Proposition}: The spherical basis under transformation of the
observer position eq.\,(\ref{gh2b}) transforms according to eq.\,(\ref{gh14}).


\subsection{Application to the spherical manifolds 
$N2 \equiv L(8,3)$, $L(8,1)$, and $N3$.}
\label{subsec:appl}


Using eq.\,(\ref{Eq:generator_N2_L83_v2}) it is easy to construct the set of polynomials $\psi(j,m_1,m_2)(u)$, see eq.\,(\ref{Eq:wigner_basis}),
which form the basis of harmonic analysis on the manifold $N2$: 
To have a Wigner polynomial 
\begin{equation}
\label{Eq:D_function_rho_alpha_epsilon}
D^j_{-m_1,m_2}(u)
=\exp\left[-\hbox{i}\,(\alpha + \epsilon)\, m_1\right]
d^j_{-m_1,m_2}(2 \rho)
\exp\left[\hbox{i}\,(\alpha - \epsilon)\, m_2\right]
\hspace{10pt}
\end{equation}
in the coordinates $u=u(\rho,\alpha,\epsilon)$ invariant under 
the action of the generator (\ref{Eq:action_generator_N2_L83_v2})
of $H=C_8$ requires
\begin{equation}
\label{e8}
\psi^{N2}(j,m_1,m_2)(u) \; =\; \psi(j,m_1,m_2)(u):\: 
2\,m_1+\,m_2 \equiv 0\;  {\rm mod}\; 4
\hspace{10pt}.
\end{equation}
In a $(m_1,m_2)$-lattice on the plane (see fig.\,5 in \cite{KR10B}), 
we can choose the sublattice with lattice basis vectors
\begin{equation}
 \label{e8b}
\vec a_1=(-1,2)
\hspace{10pt} , \hspace{10pt}
\vec a_2=(1,2)
\hspace{10pt}.
\end{equation}
Then any sublattice point obeys eq.\,(\ref{e8}),
and the harmonic basis for $N2$ consists of the towers of
polynomials $\psi(j,m_1,m_2)(u)$ over this sublattice point with
$j=j_0+\nu,\nu=0,1,2,,.., j_0={\rm Max} (|m_1|,|m_2|)$.
In general, similar eigenmodes on the lens spaces are reported 
in \cite{Lachieze-Rey_2004}
and equivalent sets of eigenmodes on the lens spaces in 
\cite{Lehoucq_et_al_2002,Lehoucq_et_al_2003}. 

On the lens space $L(8,1)$,
the action of the generator (\ref{Eq:action_generator_L81})
leads to the eigenmodes in the Wigner basis
\begin{equation}
\label{Eq:eigenfunction_L81}
\psi^{L(8,1)}(j,m_1,m_2)(u) \;=\; \psi(j,m_1,m_2)(u):\:
m_1\equiv 0 \;{\rm mod}\; 4 
\hspace{10pt}.
\end{equation}

The invariant eigenfunctions of the Laplace-Beltrami operator on $N3$ in 
the Wigner basis are determined by (\ref{Eq:generators_group_N3}) as
\begin{eqnarray}
\label{Eq:eigenfunction_N3}
\psi^{N3}(j,m_1,m_2)(u) & & \\
& & \nonumber \hspace*{-90pt} = \;
\left\{\begin{array}{ll}
\frac{1}{\sqrt{2}}\left(\psi(j,m_1,m_2)(u)+(-1)^{m_1}\,\psi(j,-m_1,m_2)(u)\right) &
: \; j \;\hbox{even}, m_1>0  \\
\psi(j,m_1,m_2)(u) &
: \; j \;\hbox{even}, m_1=0 \\
\frac{1}{\sqrt{2}}\left(\psi(j,m_1,m_2)(u)-(-1)^{m_1}\,\psi(j,-m_1,m_2)(u)\right) &
: \; j \;\hbox{odd}, m_1>0
\end{array}\right.
\end{eqnarray}
where $j\in \{0,2,3,4,...\}$, $m_2 \in {\mathbb Z}$,
$m_1 \in {\mathbb N}_0$, $m_1\equiv 0\;\hbox{mod}\;2$, 
and $m_1,\left|m_2\right|\le j$.
A similar result is stated in \cite{Lachieze-Rey_2004}
and an equivalent set of the eigenmodes in 
\cite{Lehoucq_et_al_2002,Lehoucq_et_al_2003}.

Under the transformation (\ref{Eq:trafo_coordinate})
to arbitrary new coordinates,
the relation $\psi '^{N3}(j,m_1,m_2)(\tilde u')=\psi^{N3}(j,m_1,m_2)(\tilde u)$
holds, because the eigenfunctions of the Laplace-Beltrami operator
are scalar functions.
For this reason we get in the case of $N3$ for $\tilde u =u q^{-1}$
\begin{eqnarray}
\label{Eq:eigenfunction_N3_expand}
\psi '^{N3}(j,m_1,m_2)(u) &=&  \psi^{N3}(j,m_1,m_2)(u\,q^{-1})\nonumber\\ 
&=& \sum_{\tilde{m}_2=-j}^{j}\psi^{N3}(j,m_1,\tilde{m}_2)(u)\,
D^j_{\tilde{m}_2,m_2}(q^{-1})
\end{eqnarray}
for all allowed values of $m_1$ and $m_2$.
In general such an expansion is possible for all eigenfunctions on
homogeneous spherical manifolds using the Wigner basis. 
Since for every $\left|m_2 \right|\le j$ an eigenfunction on $N3$ exists
we can choose a new equivalent basis of eigenfunctions
in the coordinates $u$ which is given by
\begin{eqnarray}
\label{Eq:eigenfunction_N3_new}
\tilde{\psi}'^{N3}(j,m_1,m_2)(u) &=& 
\sum_{\tilde{m}_2=-j}^{j}\psi '^{N3}(j,m_1,\tilde{m}_2)(u)\,
D^j_{\tilde{m}_2,m_2}(q)  \nonumber\\ 
&=& 
\sum_{\tilde{m}_2=-j}^{j}\psi^{N3}(j,m_1,\tilde{m}_2)(u\,q^{-1})\,
D^j_{\tilde{m}_2,m_2}(q)  \nonumber\\
&=& \psi^{N3}(j,m_1,m_2)(u)
\hspace{10pt}.
\end{eqnarray}
Here the transformation (\ref{Eq:eigenfunction_N3_expand}) and
$\sum_{\tilde{m}_2=-j}^{j}\,D^j_{m_1,\tilde{m}_2}(q^{-1})\,D^j_{\tilde{m}_2,m_2}(q)
=\delta_{m_1,m_2}$ are used.
The last step requires that there are no restrictions on $m_2$.
A restriction on $m_2$ would lead to observer dependent eigenfunctions.
The above calculation shows that we can choose for all coordinate shifts
$q^{-1}$ of the observer the same basis of eigenfunctions.
For this reason the CMB properties do not depend on the
position of the observer in a homogeneous manifold.

In the case of the eigenfunctions (\ref{e8}) on the other manifold $N2$,
not all values of $\left|m_2\right|\le j$ are allowed.
For this reason we cannot make an analog choice of the eigenfunctions
as on $N3$, and hence the analysis on $N2$ shows a dependence of the observer.
Expanding the eigenfunctions (\ref{e8}) on $N2$ with respect to the
basis (\ref{Eq:wigner_basis}) we get
\begin{eqnarray}
\label{Eq:eigenfunction_N2_exp}
\psi '^{N2}(j,m_1,m_2)(u) &=&  \psi^{N2}(j,m_1,m_2)(u\,q^{-1})\nonumber\\ 
&=&
\sum_{\tilde{m}_2=-j}^{j}\,\psi(j,m_1,\tilde{m}_2)(u)\,D^j_{\tilde{m}_2,m_2}(q^{-1})
\hspace{10pt}
\end{eqnarray}
with the restriction $2\,m_1+m_2 \equiv 0\;  {\rm mod}\; 4$.
This restriction on $m_2$ prohibits a new basis analogous to
(\ref{Eq:eigenfunction_N3_new}).
Here $q^{-1}$ gives the position of the observer with respect to the 
coordinate system where the Voronoi domain is given as a spherical lens,
see fig.\,\ref{fig:fundamental_cell_N2_vs_L83}.

Using eq.~(\ref{x2}) we can transform the eigenfunctions into an expansion
with respect to the spherical basis $\psi(j l m)(u)$:
\begin{eqnarray}
\label{Eq:eigenfunction_N2_exp_sph}
\psi'^{N2}(j,m_1,m_2)(u)
\nonumber &=&
\sum_{l=0}^{2j}\sum_{m=-l}^{l} \xi^{j,\rho(m_1,m_2)}_{lm}(N2;q)\,
\psi(j,l,m)(u)\;\;,\\
\xi^{j,\rho(m_1,m_2)}_{lm}(N2,q)
&=& \langle jm_1j\tilde{m}_2|lm\rangle \,D^j_{\tilde{m}_2,m_2}(q^{-1})
\hspace{10pt} 
\end{eqnarray}
where  $2\,m_1+m_2 \equiv 0\;  {\rm mod}\; 4$, $m_1+\tilde{m}_2 = m$,
and $1 \le \rho(m_1,m_2)\le r^{N2}(\beta)$ counts the 
multiplicity $r^{N2}(\beta)$ of the eigenvalue $E_j$ of the
Laplace-Beltrami operator on $N2$ for $j \in {\mathbb N}_0$.
The multiplicity for $1\le j \le 4$ is given in table\,\ref{table:multi}.
The expansion\,(\ref{Eq:eigenfunction_N2_exp_sph}) is convenient for 
the following applications.

On the homogeneous manifold $N3$
we can get a similar expansion of the eigenfunctions
\begin{eqnarray}
\label{Eq:eigenfunction_N3_exp_sph}
\psi'^{N3}(j,m_1,m_2)(u)
\nonumber &=&
\sum_{l=0}^{2j}\sum_{m=-l}^{l} \xi^{j,\rho(m_1,m_2)}_{lm}(N3)\,\psi(j, l, m)(u)\\
\xi^{j,\rho(m_1,m_2)}_{lm}(N3) & & \\ \nonumber
&& \hspace{-80pt} = \; \left\{\begin{array}{lcl}
\frac{1}{\sqrt{2}}\left(\langle jm_1jm_2|lm\rangle
+\langle j-m_1jm_2|lm\rangle \right)
& : & j \;\hbox{even}, m_1>0 \\
\langle j0jm_2|lm\rangle
& : & j \;\hbox{even}, m_1=0 \\
\frac{1}{\sqrt{2}}\left(\langle j-m_1jm_2|lm\rangle -\langle jm_1jm_2|lm \rangle \right)
& : & j \;\hbox{odd}, m_1>0 
\end{array} \right.
\hspace{10pt}
\end{eqnarray}
where $m_1 \equiv 0\;  {\rm mod}\; 2$, $0 \le m_1 \le j$,
$\left|m_2\right|\le j$, and $j \in \{0,2,3,4,...\}$.
As shown above we can choose the same expansion for all positions 
of the observer in this case.
The calculation of the ensemble average of the angular power spectrum
(\ref{Eq:Cl_ensemble_general}) 
or the 2-point correlation function (\ref{Eq:C_theta}) requires the evaluation of the
following sums.
For the homogeneous $N3$ space, using eq.\,(3.5.15) in \cite{ED64},
one obtains
\begin{eqnarray}
\label{Eq:sumrule_N3}
\nonumber
\frac{1}{2l+1}\sum_{m=-l}^{l}\sum_{m_1}\sum_{m_2=-j}^{j}
\left|\xi^{j,\rho(m_1,m_2)}_{lm}(N3)\right|^2 
\hspace{-150pt} & & \\
\nonumber 
& &= \frac{1}{2l+1}\sum_{m=-l}^{l}\sum_{m_1}\sum_{m_2=-j}^{j}
\langle jm_1jm_2|lm\rangle^2 \\
\nonumber 
& &= \frac{1}{2j+1}\sum_{m=-l}^{l}\sum_{m_1}\sum_{m_2=-j}^{j}
\langle j-m_2lm|jm_1\rangle^2 \\
& &=  \frac{1}{2j+1}\sum_{m_1}1
\; = \; \frac{r^{N3}(\beta)}{\beta^2}
\hspace{10pt},
\end{eqnarray}
and for the inhomogeneous $N2$ space with its restrictions on $m_2$
$(2m_1+m_2=0 \hbox{ mod } 4)$
\begin{eqnarray}
\label{Eq:quadratic_sum_N2}
\nonumber
\frac{1}{2l+1}\sum_{m=-l}^{l}\sum_{m_1,m_2}
\left|\xi^{j,\rho(m_1,m_2)}_{lm}(N2;q)\right|^2
\hspace{-150pt} & & \\
\nonumber
& &=\frac{1}{2l+1}\sum_{m=-l}^{l}\sum_{m_1,m_2}
\left|\langle jm_1j\tilde{m}_2|lm\rangle\,D^j_{\tilde{m}_2,m_2}(q^{-1})\right|^2 \\
& &=\frac{1}{2l+1}\sum_{m=-l}^{l}\sum_{m_1,m_2}
\left|\langle jm_1j\tilde{m}_2|lm\rangle\,d^j_{\tilde{m}_2,m_2}(-2 \rho)\right|^2 
\hspace{10pt}.
\end{eqnarray}
In the derivation of eq.\,(\ref{Eq:quadratic_sum_N2}) we have used
eq.~(\ref{Eq:D_function_rho_alpha_epsilon})
where the coordinates of the observer position on the manifold are 
parameterised by
eq.\,(\ref{Eq:coordinate_q_rho_alpha_epsilon}). 
The $d$-functions are computed using
the algorithm described in \cite{Risbo_1996}.
The important point is
that eq.\,(\ref{Eq:sumrule_N3}) is independent of the observer position,
whereas eq.\,(\ref{Eq:quadratic_sum_N2}) has an explicit $\rho$ dependence.
Eq.\,(\ref{Eq:sumrule_N3}) applies on all homogeneous spherical 
manifolds ${\cal M}$ using the corresponding multiplicity $r^{\cal M}(\beta)$
of the eigenmodes \cite{AU05,Gundermann_2005,AU05b,Bellon_2006,Lustig_2007}.
Thus this equation is also true for the manifold $L(8,1)$.

The following transformation shows
that the observers at $\rho$ and at $\pi/2-\rho$ see the same
CMB anisotropies in the statistical sense.
Substituting in eq.\,(\ref{Eq:quadratic_sum_N2}) $-2\rho$ by $\pi+2\rho$
one gets
\begin{eqnarray}
\label{Eq:quadratic_sum_N2_sym}
\nonumber &  &\frac{1}{2l+1}\sum_{m=-l}^{l}\sum_{m_1,m'_2}
\left|\langle jm_1j\tilde{m}_2|lm\rangle\,
d^j_{\tilde{m}_2,m'_2}(\pi+2 \rho)\right|^2  \\
\nonumber & & = \frac{1}{2l+1}\sum_{m=-l}^{l}\sum_{m_1,m'_2}
\left|\langle jm_1j\tilde{m}_2|lm\rangle\,
\left(-1\right)^{(j-\tilde{m}_2)}\,d^j_{-\tilde{m}_2,m'_2}(-2 \rho)\right|^2 \\
\nonumber& & = \frac{1}{2l+1}\sum_{m=-l}^{l}\sum_{m_1,m_2}
\left|\langle jm_1j-\tilde{m}_2|lm\rangle\,
d^j_{\tilde{m}_2,m_2}(-2 \rho)\right|^2 \\
\nonumber& & = \frac{1}{2l+1}\sum_{m=-l}^{l}\sum_{m_1,m_2}
\left|\langle j-m_1j\tilde{m}_2|l-m\rangle\,
d^j_{\tilde{m}_2,m_2}(-2 \rho)\right|^2\\
& & = \frac{1}{2l+1}\sum_{m=-l}^{l}\sum_{m_1,m_2}
\left|\langle jm_1j\tilde{m}_2|lm\rangle\,
d^j_{\tilde{m}_2,m_2}(-2 \rho)\right|^2
\hspace{10pt}
\end{eqnarray} 
where eqs.\,(3.5.17) and (4.2.4) in \cite{ED64} are used
and that there exists to every eigenfunction with the numbers $(m_1,m_2)$ 
also an eigenfunction with the numbers $(-m_1,m_2)$ on the manifold $N2$,
see eq.\,(\ref{e8b}). 
Thus, the CMB analysis of the space $N2$ can be restricted to the
interval $\rho \in [0,\pi/4]$.

\begin{table}
\begin{center}
\begin{tabular}{|c|c|c|c|c|}
\hline
$\beta \; = \; 2j+1$ & $r^{L(8,1)}(\beta)$ & $r^{N2}(\beta)$ & $r^{N3}(\beta)$ 
& $r^{{\mathbb S}^3}(\beta)/8$ \\
\hline
3 & 3 & 1 & 0 & 1.125 \\
\hline
4 & 0 & 0 & 0 & 2 \\
\hline
5 & 5 & 7 & 10 & 3.125 \\
\hline
6 & 0 & 0 & 0 & 4.5 \\
\hline
7 & 7 & 11 & 7 & 6.125 \\
\hline
8 & 0 & 0 & 0 & 8 \\
\hline
9 & 27 & 23 & 27 & 10.125  \\
\hline
\end{tabular}
\end{center}
\caption{\label{table:multi} 
The multiplicity $r^{\cal M}(\beta)$ of the eigenmodes for 
the wave numbers $\beta$ from 3 to 9 on the manifolds ${\cal M} = L(8,1)$, 
$N2\equiv L(8,3)$, and $N3$ is specified.
In the case of the 3-sphere the effective multiplicity is given
which is the multiplicity $r^{{\mathbb S}^3}(\beta)= \beta^2$ 
divided by the order of the group $N_{L(8,1)}= N_{N2}= N_{N3}=8$.
Analytical formulae for the multiplicity of the eigenmodes are known
for homogeneous spherical manifolds \cite{Ikeda_1995,Weeks_2006} and for a specific class of 
the inhomogeneous spherical manifolds \cite{Lustig_2007}
but not for the manifold $N2$.
}
\end{table}

\section{Observer dependence of the temperature 2-point correlation function
of the CMB radiation.}
\label{sec:cmb_correlation_observ}

The temperature correlations of the CMB sky with respect to their
separation angle $\vartheta$ are an important diagnostic tool.
The correlations at large angles $\vartheta$,
where the topological signature is expected,
are most clearly revealed by the temperature 2-point correlation function
$C(\vartheta)$ which is defined as
\begin{equation}
\label{Eq:C_theta}
C(\vartheta) \; := \; \left< \delta T(\hat n) \delta T(\hat n')\right>
\hspace{10pt} \hbox{with} \hspace{10pt}
\hat n \cdot \hat n' = \cos\vartheta
\hspace{10pt} ,
\end{equation}
where $\delta T(\hat n)$ is the temperature fluctuation in
the direction of the unit vector $\hat n$.
The 2-point correlation function $C(\vartheta)$ is related
to the multipole moments $C_l$ by
\begin{equation}
\label{Eq:C_theta_Cl}
C(\vartheta) \; = \; 
\sum_l\,\frac{2l+1}{4\pi}\,C_l\,P_l\left(\cos\vartheta\right)
\hspace{10pt}.
\end{equation}
The ensemble average of $C_l$ can be expressed for a spherical
manifold ${\cal M}$ by the expansion coefficients
$\xi^{\beta,\rho}_{lm}({\cal M};q)$ discussed in the previous section
\begin{eqnarray}
\label{Eq:Cl_ensemble_general}
C_l & := &
\frac{1}{2l+1}\sum_{m=-l}^l\left\langle\left|a_{lm}\right|^2\right\rangle
\\ & = &\nonumber
\sum_{\beta}\frac{ T_l^2(\beta) \; P(\beta)}{2l+1}\sum_{m=-l}^{l}\sum_{\rho}\left|\xi^{\beta,\rho}_{lm}({\cal M};q)\right|^2
\hspace{10pt} ,
\end{eqnarray}
with the initial power spectrum  $P(\beta)\sim 1/(E_{\beta}\,\beta^{2-n_s})$
where $E_{\beta}=\beta^2-1$ are the allowed eigenvalues of the Laplace-Beltrami
operator on the considered spherical manifold ${\cal M}$, 
$\beta=2j+1$, and $n_s$ is the spectral index.
$T_l(k)$ is the transfer function containing the full Boltzmann physics,
e.\,g.\ the ordinary and the integrated Sachs-Wolfe effect, 
the Doppler contribution, the Silk damping and the reionization 
are taken into account.
Using the expression (\ref{Eq:quadratic_sum_N2}) for the 
expansion coefficients $\xi^{\beta,\rho}_{lm}({\cal M};q)$ 
we get for the ensemble average of $C_l$ on $N2$
\begin{equation}
\label{Eq:Cl_ensemble_N2}
C_l \; = \;
\sum_{\beta}\frac{ T_l^2(\beta) \; P(\beta)}{2l+1}\sum_{m=-l}^{l}\sum_{m_1,m_2}
\left|\langle jm_1j\tilde{m}_2|lm\rangle\,d^j_{\tilde{m}_2,m_2}(-2\rho)\right|^2
\end{equation}
which depends only on the distance $\rho$.
Therefore, also the ensemble average of the 2-point correlation function 
on $N2$ depends only on $\rho$. 
Since the sum over the expansion coefficient in eq.\,(\ref{Eq:Cl_ensemble_N2})
fulfils the symmetry~(\ref{Eq:quadratic_sum_N2_sym}) the interval of $\rho$
can be restricted to $\rho \in [0,\frac{\pi}{4}]$. 

In contrast, using the sum rule (\ref{Eq:sumrule_N3}),
the ensemble average of $C_l$ on $N3$ is given by
\begin{equation}
\label{Eq:Cl_ensemble_N3}
C_l \; = \;
\sum_{\beta}\,T_l^2(\beta) \; P(\beta)\,\frac{r^{N3}(\beta)}{\beta^2}
\end{equation}
which does not depend on the position of the observer.
The multiplicity $r^{N3}(\beta)$ restricts the sum over $\beta$ to
$\beta \geq 5$ for $N3$.
The relation (\ref{Eq:Cl_ensemble_N3}) is valid for
homogeneous manifolds.
Thus, this equation also holds for the manifold $L(8,1)$ using the 
corresponding multiplicity of the eigenmodes on $L(8,1)$.

To speed up the calculations of the ensemble average of $C_l$
on all three manifolds ${\cal M}=$ L(8,3)($\equiv N2$), $N3$ and $L(8,1)$ 
we have used for $\beta>50$ the spectrum of the projective space 
${\mathbb P}^3$ divided by $V_{\cal M}/V_{{\mathbb P}^3}=4$.
We have checked numerically that this is good approximation 
of the exact result.
This approximation can be used in a similar way for all 
manifolds which tessellate the 3-sphere under a group of 
deck transformations of even order.

\begin{figure}
\begin{center}
\vspace*{-45pt}
\hspace*{10pt}
\includegraphics[width=0.9\textwidth]{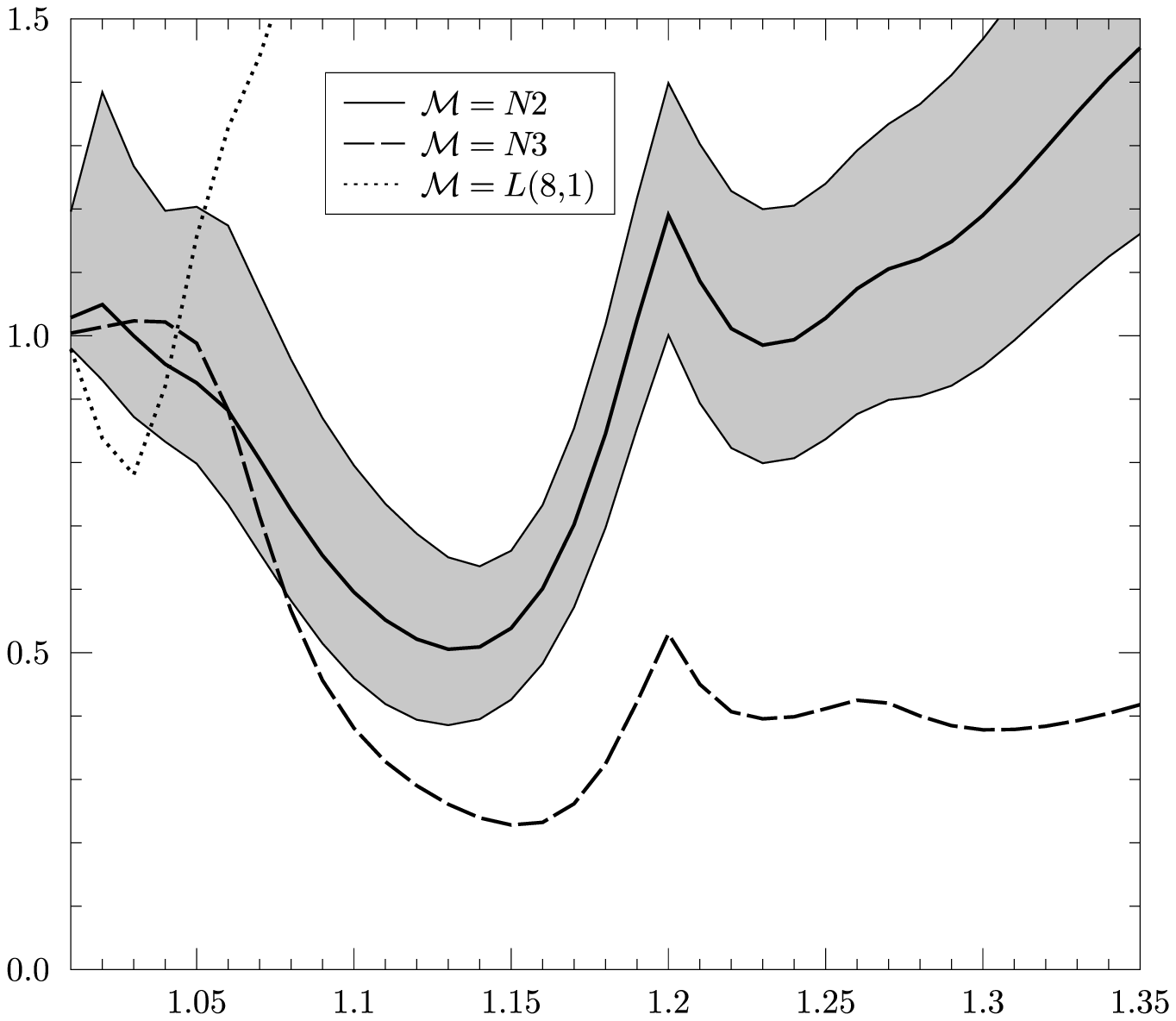}
\put(-90,35){$\Omega_{\hbox{\scriptsize tot}}$}
\put(-390,277){$\frac{S_{{\cal M}}(60^{\circ})}{S_{{\mathbb S}^3}(60^{\circ})}$}
\put(-315,147){a)}
\vspace*{-33pt}
\includegraphics[width=0.8\textwidth]{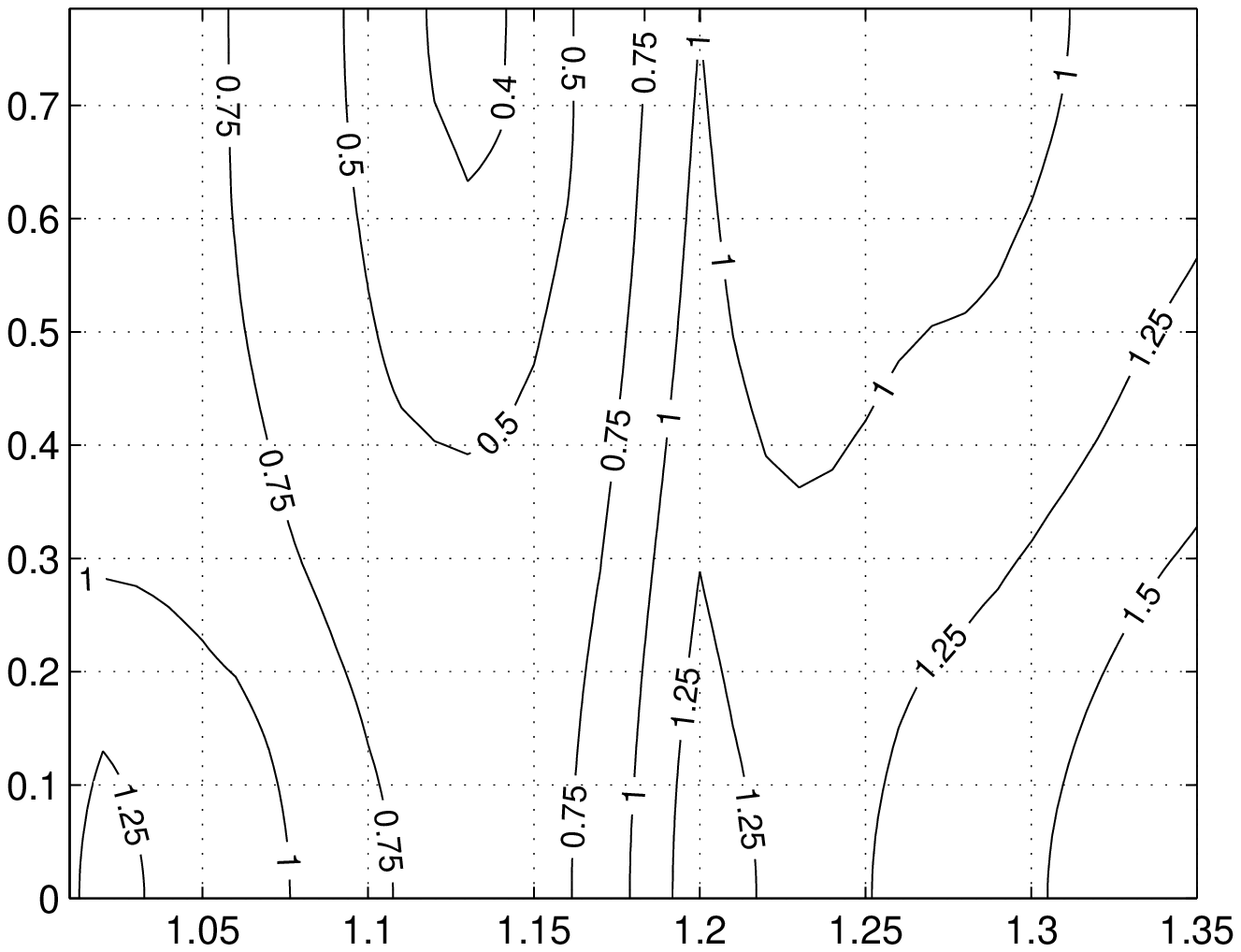}
\put(-60,10){$\Omega_{\hbox{\scriptsize tot}}$}
\put(-330,247){$\rho$}
\put(-290,147){b)}
\vspace*{-25pt}
\end{center}
\caption{\label{fig:S60}
Panel a) shows the $S_{{\cal M}}(60^{\circ})$ statistics of the manifolds
${\cal M}=L(8,1)$, $L(8,3)\equiv N2$, and $N3$ normalised to 
the $S_{{\mathbb S}^3}(60^{\circ})$ statistics of the 3-sphere ${\mathbb S}^3$ 
depending on the total density parameter $\Omega_{\hbox{\scriptsize tot}}$. 
In the case of the manifold $N2\equiv L(8,3)$ the average of the 
$S_{L(8,3)}(60^{\circ})$ statistics due to the observer dependence is displayed 
as a full line. 
The dispersion of the $S_{L(8,3)}(60^{\circ})$ statistics depending 
on the observer is given by a grey band.
In panel b) the $S_{L(8,3)}(60^{\circ})$ statistics 
normalised to the $S_{{\mathbb S}^3}(60^{\circ})$ statistics depending
on $\Omega_{\hbox{\scriptsize tot}}$ and the observer position is displayed 
where the position of the observer is characterised by $\rho$.  
}
\end{figure}

The comparison of the various correlation functions $C(\vartheta)$
is facilitated by the introduction of the $S(60^{\circ})$ statistics
\begin{eqnarray}
\label{Eq:S_60}
S(60^{\circ}):= \int_{-1}^{\cos(60^{\circ})}\hbox{d}\cos\vartheta 
\left| C(\vartheta) \right|
\hspace{10pt}.
\end{eqnarray}
It quantifies the power of the 2-point correlation function $C(\vartheta)$ 
on scales large than $\vartheta=60^{\circ}$
which is the interesting scale for topological studies. 
This scalar quantity has been introduced in \cite{Spergel_et_al_2003}
in order to describe the low power at large angular scales
which has been observed in the CMB sky.

In the following, the 2-point correlation functions are calculated for 
the cosmological parameters $\Omega_{\hbox{\scriptsize cdm}} = 0.238$,
$\Omega_{\hbox{\scriptsize bar}} = 0.0485$, $h=0.681$, and
$n_{\hbox{\scriptsize s}}= 0.961$.
The density parameter of the cosmological constant $\Omega_{\Lambda}$ is
changed according to get the desired total density
parameter $\Omega_{\hbox{\scriptsize tot}}$.

In fig.\,\ref{fig:S60}a
the $S(60^{\circ})$ statistics is presented for a wide range of
the total density parameter $\Omega_{\hbox{\scriptsize tot}}$
for the three manifolds $L(8,1)$, $L(8,3)\equiv N2$, and $N3$.
To emphasise the differences between the different topological spaces,
the $S(60^{\circ})$ statistics is normalised with respect to
the simply-connected manifold ${\mathbb S}^3$,
that is, the plot shows
$\frac{S_{{\cal M}}(60^{\circ})}{S_{{\mathbb S}^3}(60^{\circ})}$.
For these three manifolds the ensemble averages with respect to
the sky realisations are shown.
For the two homogeneous spaces $L(8,1)$ and $N3$ there is no observer
dependence and thus, the result is shown as a dotted and dashed curve,
respectively.
The inhomogeneous space $L(8,3)$ has a dependence on the observer position,
which can be parameterised by the distance $\rho$ as discussed above.
The variability of the ensemble average with respect to $\rho$ is shown
as the grey band.
The average over the interval $\rho \in [0,\pi/4]$ is plotted as the
full curve.
For most values of $\Omega_{\hbox{\scriptsize tot}}<1.2$ the power in the
large scale correlation is indeed lower than for the simply-connected
manifold ${\mathbb S}^3$ since the values are smaller than one.
An even stronger suppression of large scale power is, however, revealed
by the homogeneous space $N3$ for $\Omega_{\hbox{\scriptsize tot}}>1.07$.

The grey band in fig.\,\ref{fig:S60}a does not betray which values
of the parameter $\rho$ lead to the strongest suppression of power,
that is, where the most probable observer positions occur.
This information is provided by fig.\,\ref{fig:S60}b
where for the space $L(8,3)$ the normalised power
$\frac{S_{{\cal M}}(60^{\circ})}{S_{{\mathbb S}^3}(60^{\circ})}$
is shown in dependence on $\rho$ and $\Omega_{\hbox{\scriptsize tot}}$.
Since the observer dependence is one-dimensional for the space $L(8,3)$,
the observer variability is exhaust in this topology.
Very low values of power are observed close to  
$\Omega_{\hbox{\scriptsize tot}} \simeq 1.13$ and $\rho \gtrsim 0.6$.
Comparing these values for $\rho$ with the Voronoi domains
shown in fig.\,\ref{fig:fundamental_cell_N2_vs_L83},
it is obvious that the best observer position belongs to the cubic
Voronoi domain with $\rho=\frac\pi 4$.
The observer position belonging to the lens shaped domain $(\rho=0)$
is much worse than that of the cubic domain.
The observation that more symmetrical domains suppress the large scale
CMB power better than asymmetrical domains,
is claimed in \cite{Weeks_Luminet_Riazuelo_Lehoucq_2005}
where the expression ``well proportioned'' universes is coined.
Well proportioned domains possess in all directions approximately
equal extensions, whereas oddly proportioned spaces extend in some
directions much more than in others.
The analysis of the $L(8,3)$ space demonstrates
that this rule seems to be valid even for a single manifold
provided it is inhomogeneous.
Therefore, it would be premature to exclude lens spaces, in general,
as oddly proportioned spaces with high CMB power on large angular scales.
Our analysis shows that the class of inhomogeneous spaces leads to a vast
number of models
which could in principle represent admissible models for our Universe.

\begin{figure}
\begin{center}
\hspace*{-35pt}\includegraphics[width=0.65\textwidth]{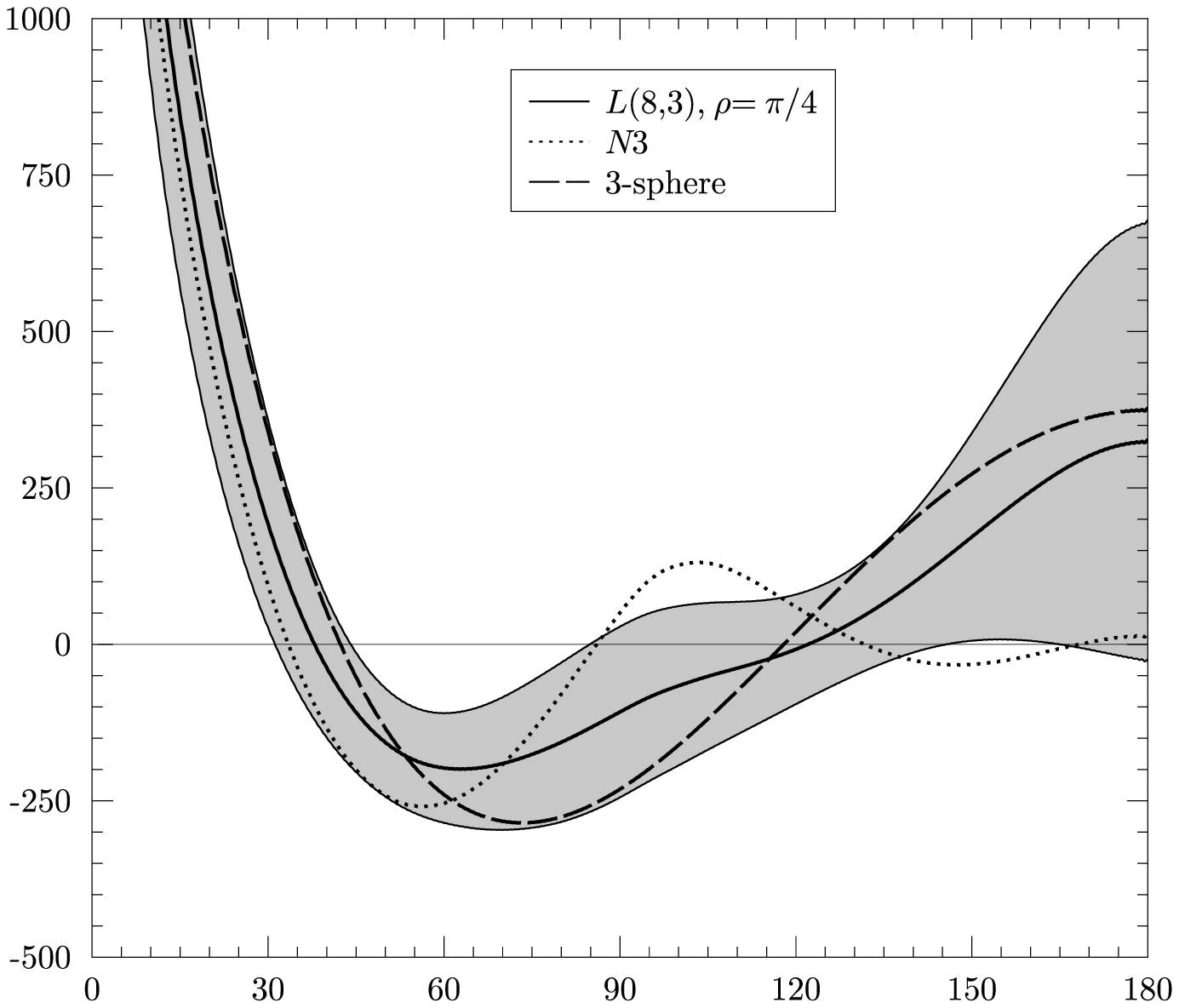}
\put(-275,200){$C(\vartheta)$}
\put(-65,27){$\vartheta$}
\put(-210,190){a)}
\hspace*{-55pt}\includegraphics[width=0.65\textwidth]{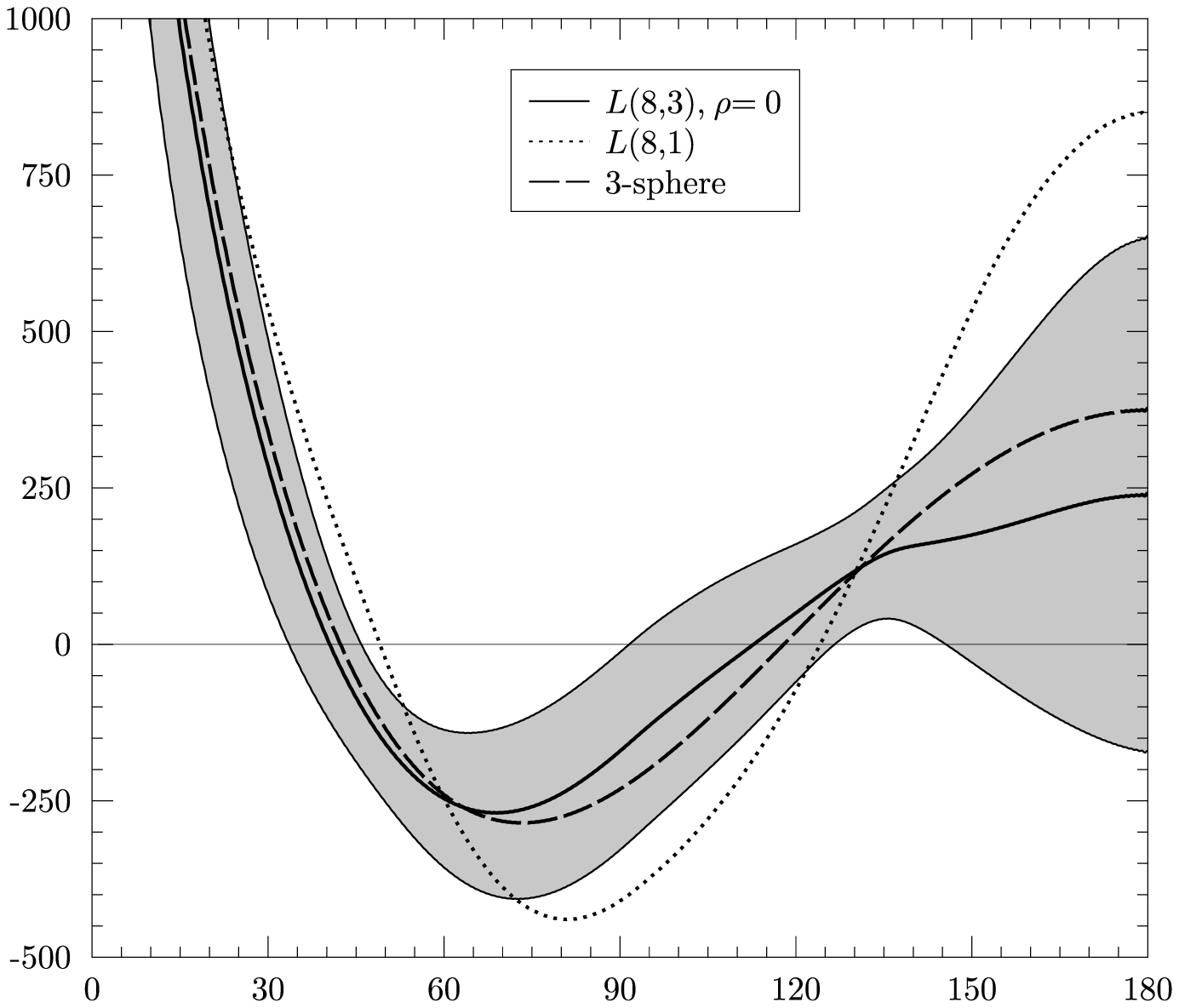}
\put(-275,200){$C(\vartheta)$}
\put(-65,27){$\vartheta$}
\put(-210,190){b)}
\vspace*{-30pt}
\end{center}
\caption{\label{fig:2-point-correlation}
In panel a) the 2-point correlation functions $C(\vartheta)$ for the manifolds 
$L(8,3)$ with the observer at $\rho=\pi/4$ ($\equiv N2$), $N3$, and 
${\mathbb S}^3$ for $\Omega_{\hbox{\scriptsize tot}}=1.13$ are displayed.
Panel b) shows $C(\vartheta)$ for the manifolds 
$L(8,3)$ with the observer at $\rho=0$ ($\equiv N2$), $L(8,1)$ and 
${\mathbb S}^3$. 
The 1-$\sigma$ standard deviation of the cosmic variance is pictured
as grey band for the inhomogeneous $L(8,3)$ space.
}
\end{figure}

Because the $S(60^{\circ})$ statistics integrates the correlation function
$C(\vartheta)$, the angular information is missing.
To reveal it, several correlation functions $C(\vartheta)$ are displayed in
fig.\,\ref{fig:2-point-correlation}.
The correlation function $C(\vartheta)$ is shown for the inhomogeneous
$L(8,3)\equiv N2$ topological space for the two observer positions
characterised by $\rho=\pi/4$ and $\rho=0$.
The first case leads to a cubic Voronoi domain
(see fig.\,\ref{fig:fundamental_cell_N2_vs_L83}d)
and the second to a lens shaped Voronoi domain
(see fig.\,\ref{fig:fundamental_cell_N2_vs_L83}a).
These are compared in figs.\,\ref{fig:2-point-correlation}a and
\ref{fig:2-point-correlation}b with the cubic $N3$ space
and the lens shaped $L(8,1)$, respectively.
For comparison, both panels also display $C(\vartheta)$
of the simply-connected ${\mathbb S}^3$ manifold.
The grey band is the cosmic variance with respect to the fixed
observer position of the $L(8,3)$ space.
Note, that the grey band in fig.\,\ref{fig:S60}a is due to the various
observer positions and does not include the cosmic variance.
In the case of the cubic Voronoi domains (fig.\,\ref{fig:2-point-correlation}a),
the correlation function $C(\vartheta)$ of $N3$ is almost contained
within the cosmic variance of the $L(8,3)$ space with $\rho=\pi/4$.
For the lens shaped Voronoi domains, the difference is more pronounced
as revealed by fig.\,\ref{fig:2-point-correlation}b.
The amplitude of $C(\vartheta)$ of the $L(8,1)$ space exceeds that
of the ${\mathbb S}^3$ manifold for most angles $\vartheta$.

\begin{figure}
\begin{center}
\hspace*{-35pt}\includegraphics[width=0.65\textwidth]{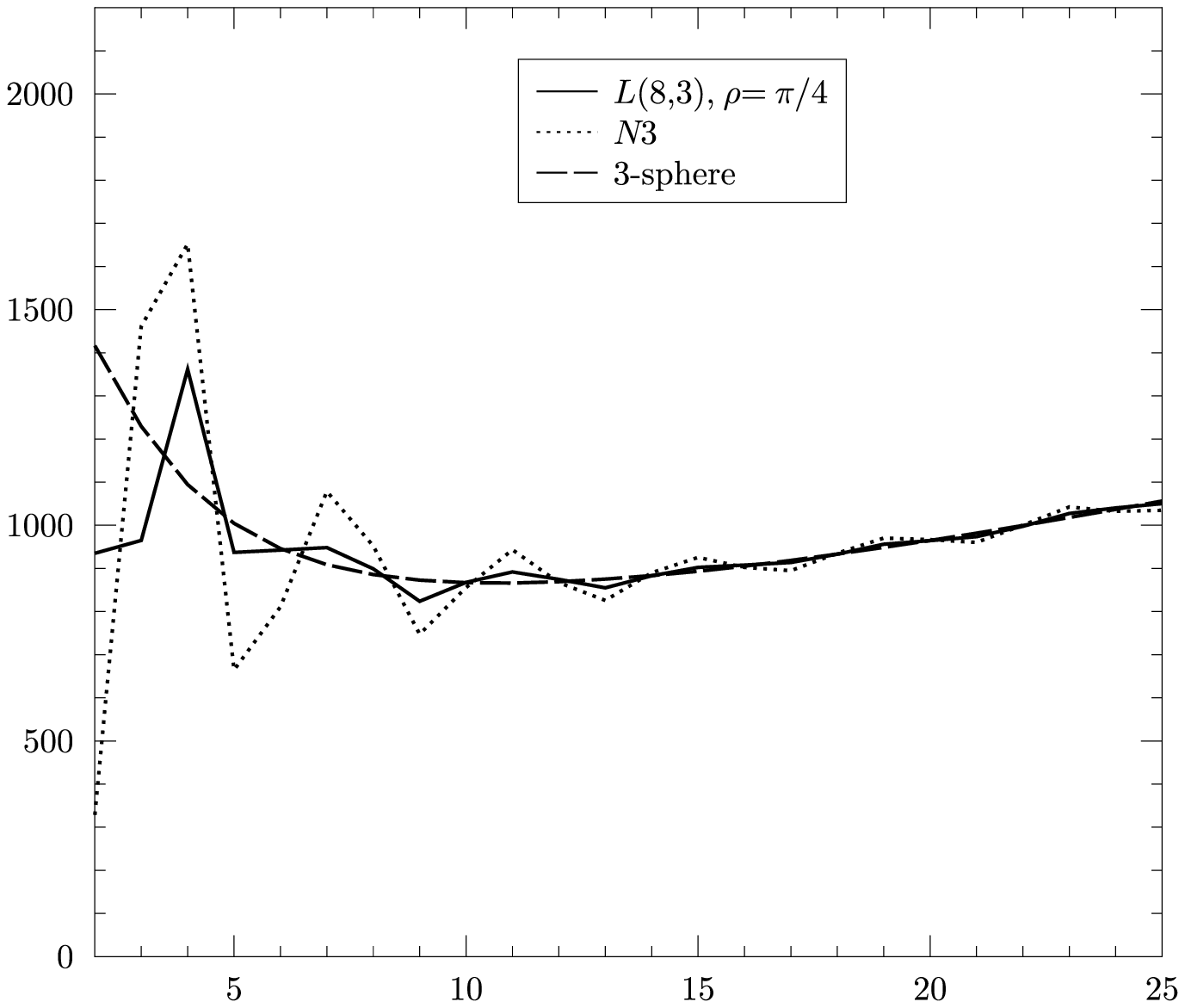}
\put(-270,180){$\delta T_{l}^{2}$}
\put(-70,27){$l$}
\put(-230,200){a)}
\hspace*{-55pt}\includegraphics[width=0.65\textwidth]{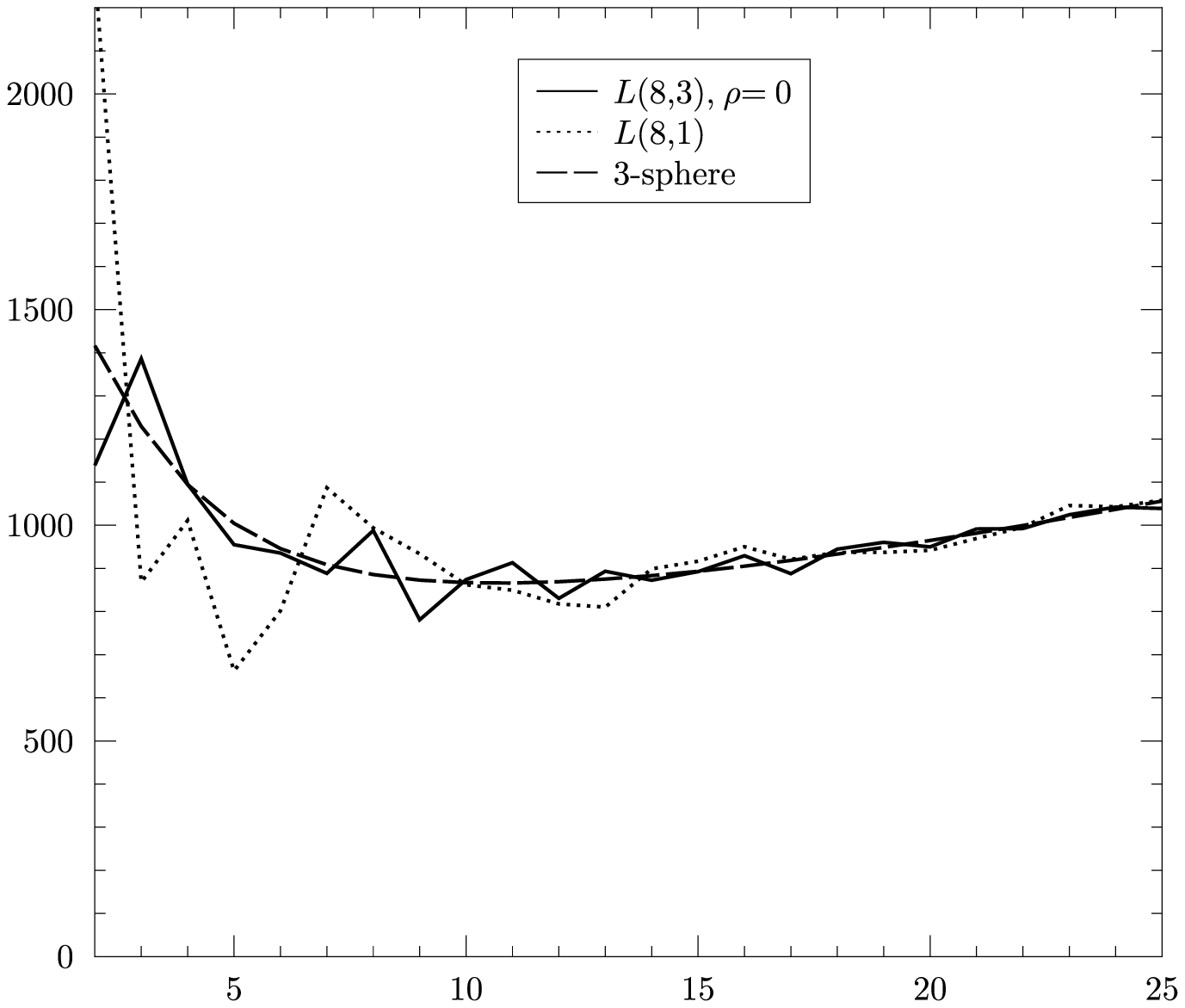}
\put(-270,180){$\delta T_{l}^{2}$}
\put(-70,27){$l$}
\put(-230,200){b)}
\end{center}\vspace*{-30pt}
\caption{\label{fig:spectrum}
The angular power spectra $\delta T_{l}^{2} = l(l+1) C_l /(2\pi)$
corresponding to the 2-point correlation functions $C(\vartheta)$
shown in fig.\,\ref{fig:2-point-correlation} are displayed.
The spectra $\delta T_{l}^{2}$ are plotted in units $[\mu\hbox{K}^2]$.
}
\end{figure}

The suppression of power at large angular scales can also be studied
by the angular power spectrum $C_l$
which is related to $C(\vartheta)$ by eq.\,(\ref{Eq:C_theta_Cl}).
The large scale suppression is revealed by small values of $C_l$
for small $l$.
For larger values of $l$, the differences due to the topological
signatures vanish.
Most interesting is the quadrupole moment $C_2$.
For the correlation functions $C(\vartheta)$
which are shown in fig.\,\ref{fig:2-point-correlation},
the corresponding angular power spectra $\delta T_l^2 = l(l+1) C_l /(2\pi)$
are plotted in fig.\,\ref{fig:spectrum}.
The quadrupole suppression of the cubic Voronoi domains
compared to the ${\mathbb S}^3$ space is obviously visible
in fig.\,\ref{fig:spectrum}a.
An extreme suppression is observed for the $N3$ space
due to the absence of the eigenmodes with $\beta < 5$,
see eq.\,(\ref{Eq:Cl_ensemble_N3}) and table \ref{table:multi}.
As fig.\,\ref{fig:spectrum}b reveals
the lens shaped Voronoi domain of $L(8,3)$ possesses a slight
suppression compared to ${\mathbb S}^3$,
but for $L(8,1)$ a strong enhancement of power is generated by
the presence of eigenmodes with $\beta=3$.
The inhomogeneous space $L(8,3)$ has also an eigenmodes with $\beta=3$,
but the multiplicity is only one
whereas it is for $L(8,1)$ threefold higher,
see table \ref{table:multi}.
The differences in the amplitudes of the quadrupole moment $C_2$ for $L(8,3)$
which are revealed by panels a) and b),
are due to the position dependent phases occurring
in eq.\,(\ref{Eq:Cl_ensemble_N2}).

The figs.\,\ref{fig:spectrum}a and \ref{fig:spectrum}b reveal that the
angular power spectrum $\delta T_l^2$ for the ${\mathbb S}^3$ space
is smooth whereas the non-trivial topologies possess strongly
fluctuating multipoles at small values of $l$.
This is caused by the erratic fluctuating multiplicities
in the case of the non-trivial topologies
and the smooth, monotonically increasing multiplicities for ${\mathbb S}^3$
(see table \ref{table:multi}).

\begin{figure}
\vspace*{-15pt}
\begin{minipage}{17.0cm}
\hspace*{-15pt}
\begin{minipage}{9.5cm}
{
\hspace*{-15pt}
\includegraphics[width=9.5cm]{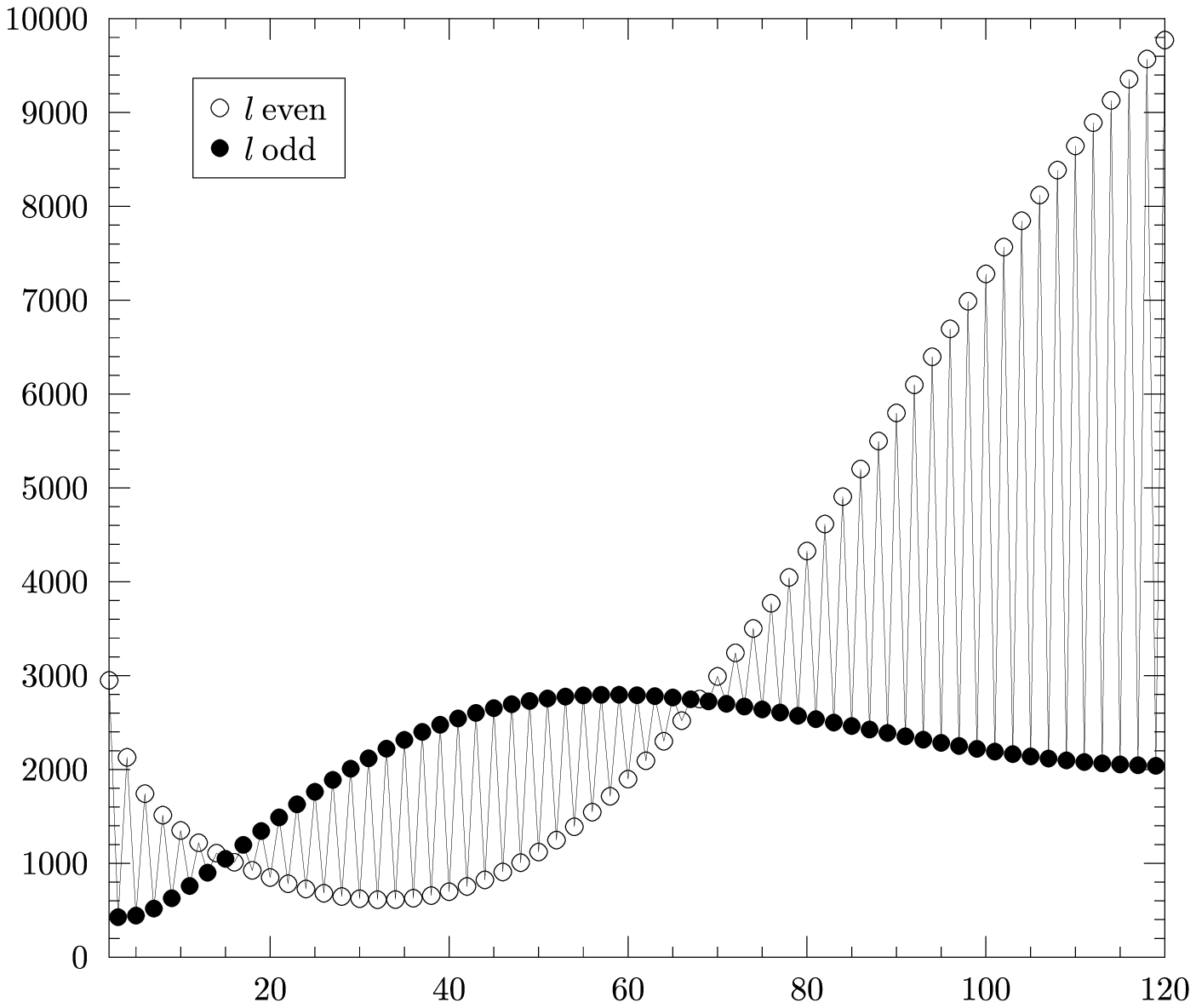}
\put(-262,190){$\delta T_{l}^{2}$}
\put(-55,20){$l$}
\put(-110,188){a) total}
\vspace*{-45pt}
}
{
\includegraphics[width=9.5cm]{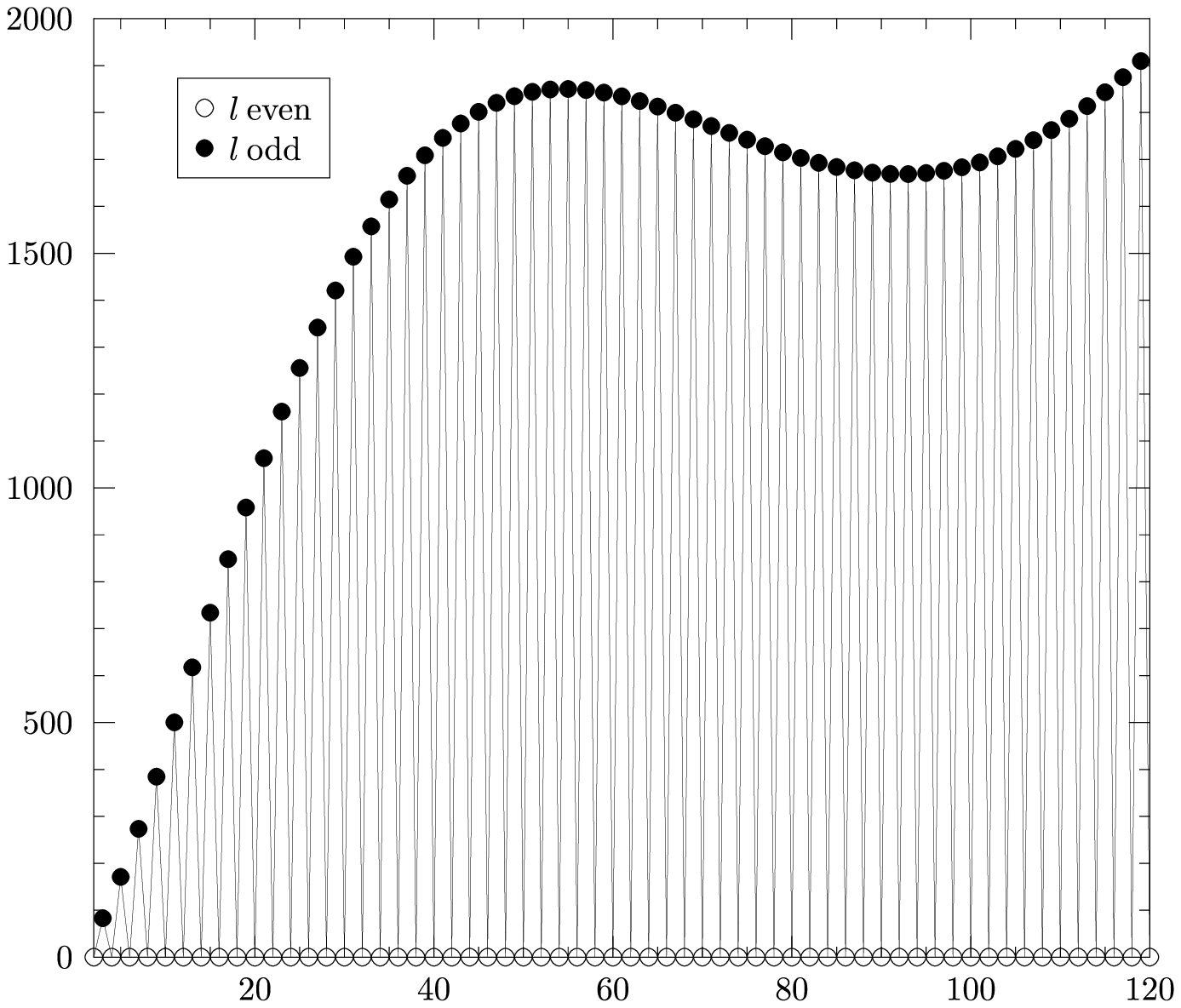}
\put(-262,190){$\delta T_{l}^{2}$}
\put(-55,20){$l$}
\put(-115,188){c) Doppler}
\vspace*{-45pt}
}
{
\includegraphics[width=9.5cm]{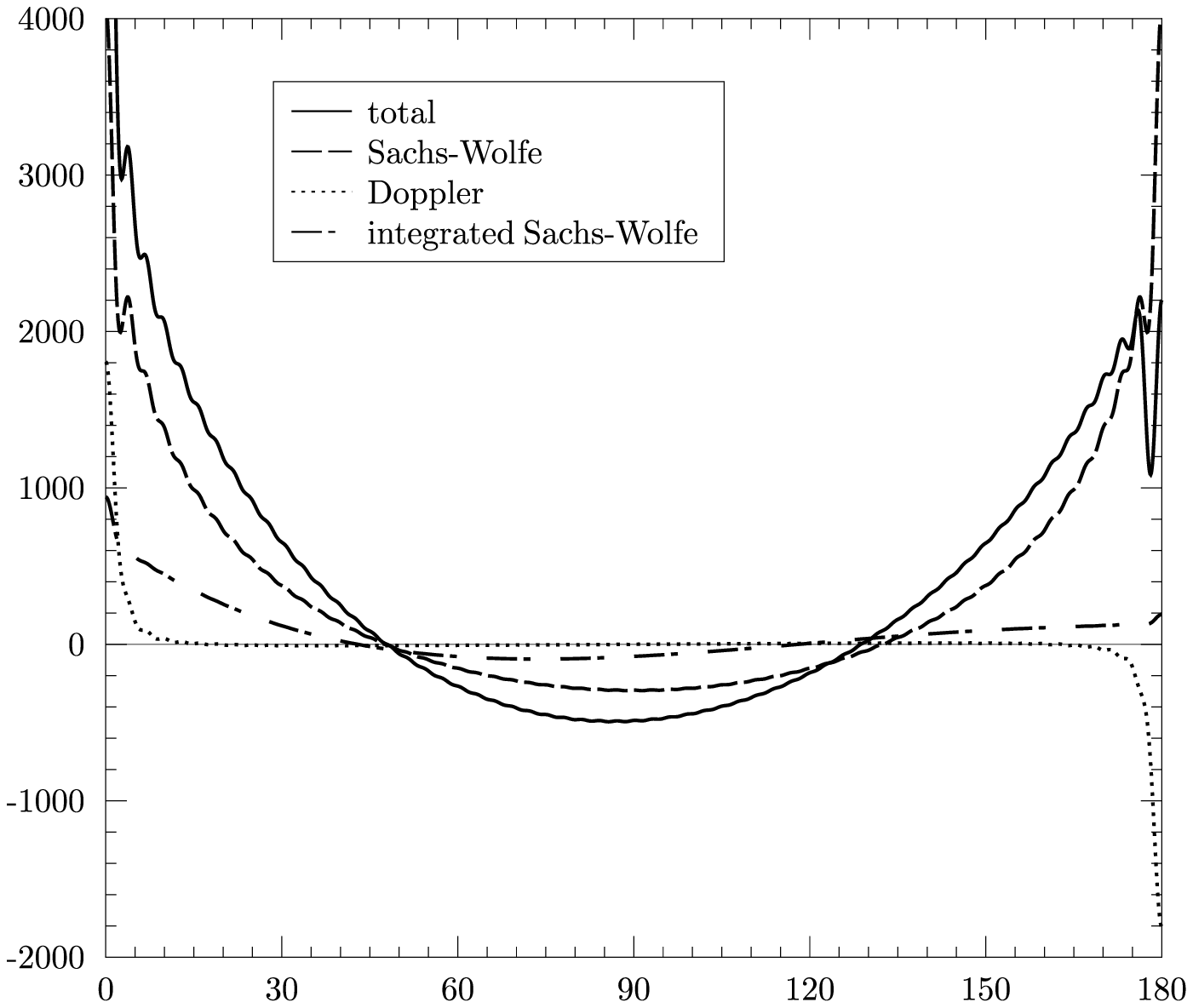}
\put(-262,188){$C(\vartheta)$}
\put(-55,20){$\vartheta$}
\put(-70,188){e)}
\vspace*{-35pt}
}
\end{minipage}
\hspace*{-45pt}
\begin{minipage}{9.5cm}
{
\includegraphics[width=9.5cm]{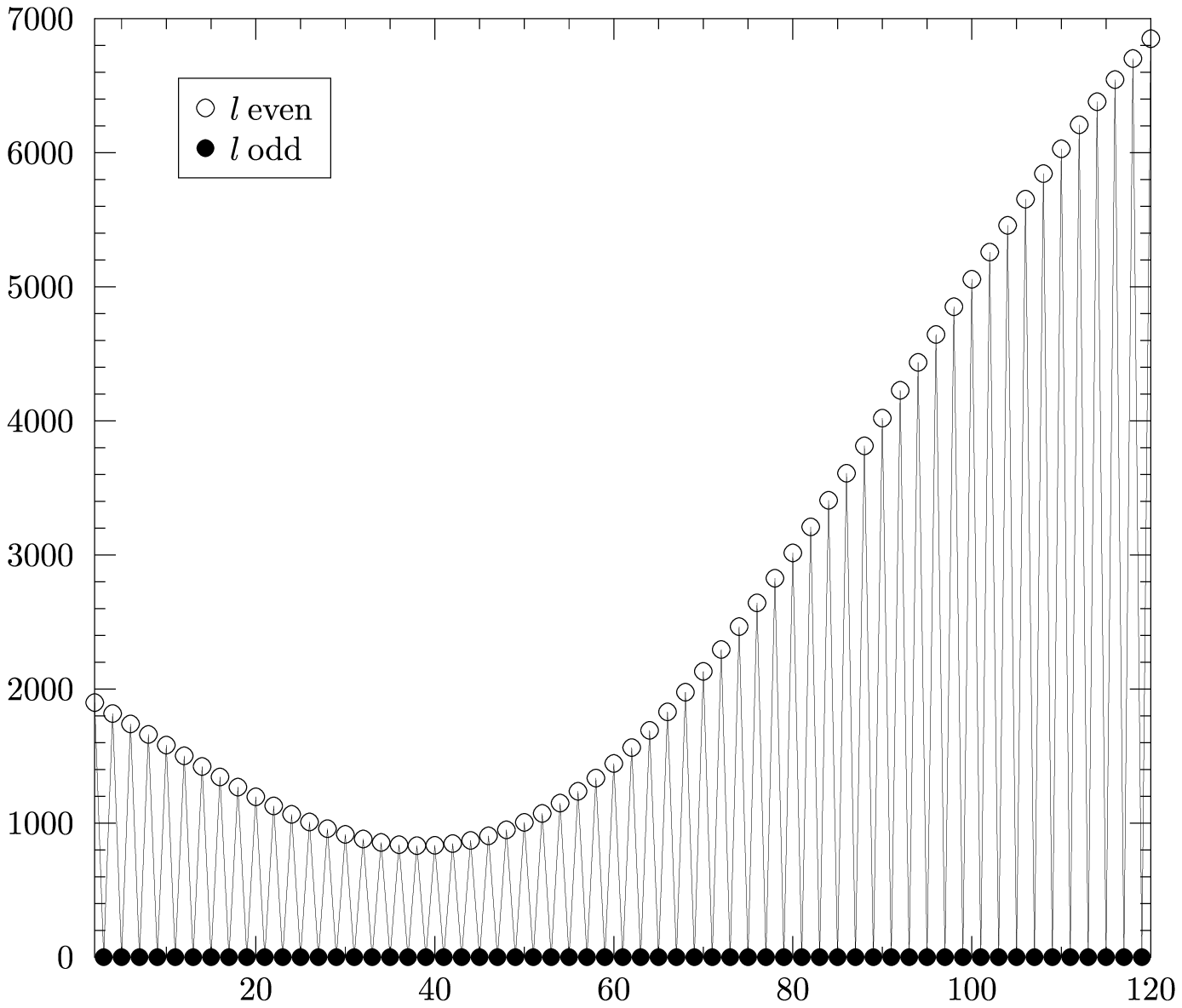}
\put(-262,190){$\delta T_{l}^{2}$}
\put(-55,20){$l$}
\put(-140,188){b) Sachs-Wolfe}
\vspace*{-45pt}
}
{
\includegraphics[width=9.5cm]{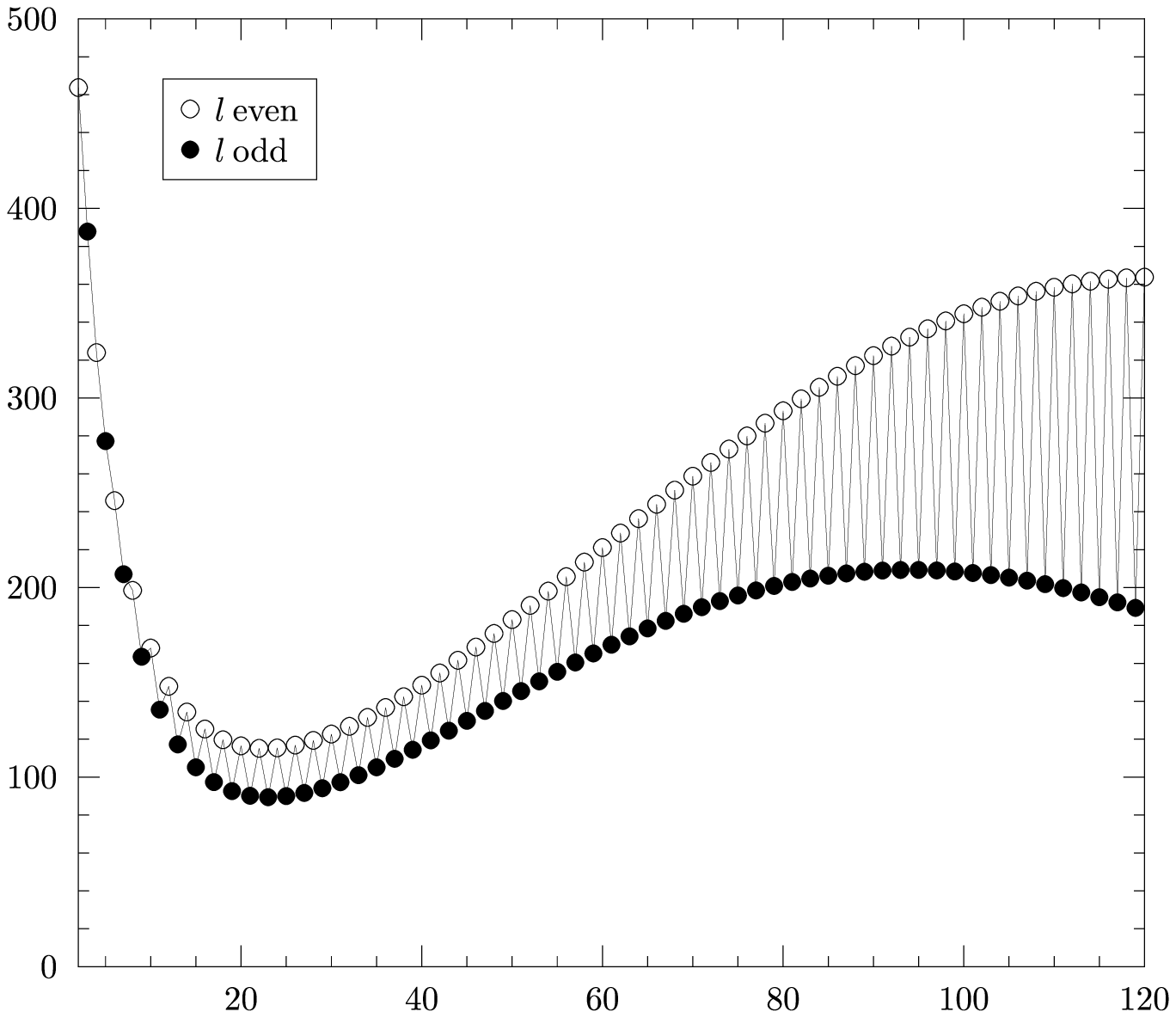}
\put(-262,190){$\delta T_{l}^{2}$}
\put(-55,20){$l$}
\put(-180,188){d) integrated Sachs-Wolfe}
\vspace*{-45pt}
}
{
\includegraphics[width=9.5cm]{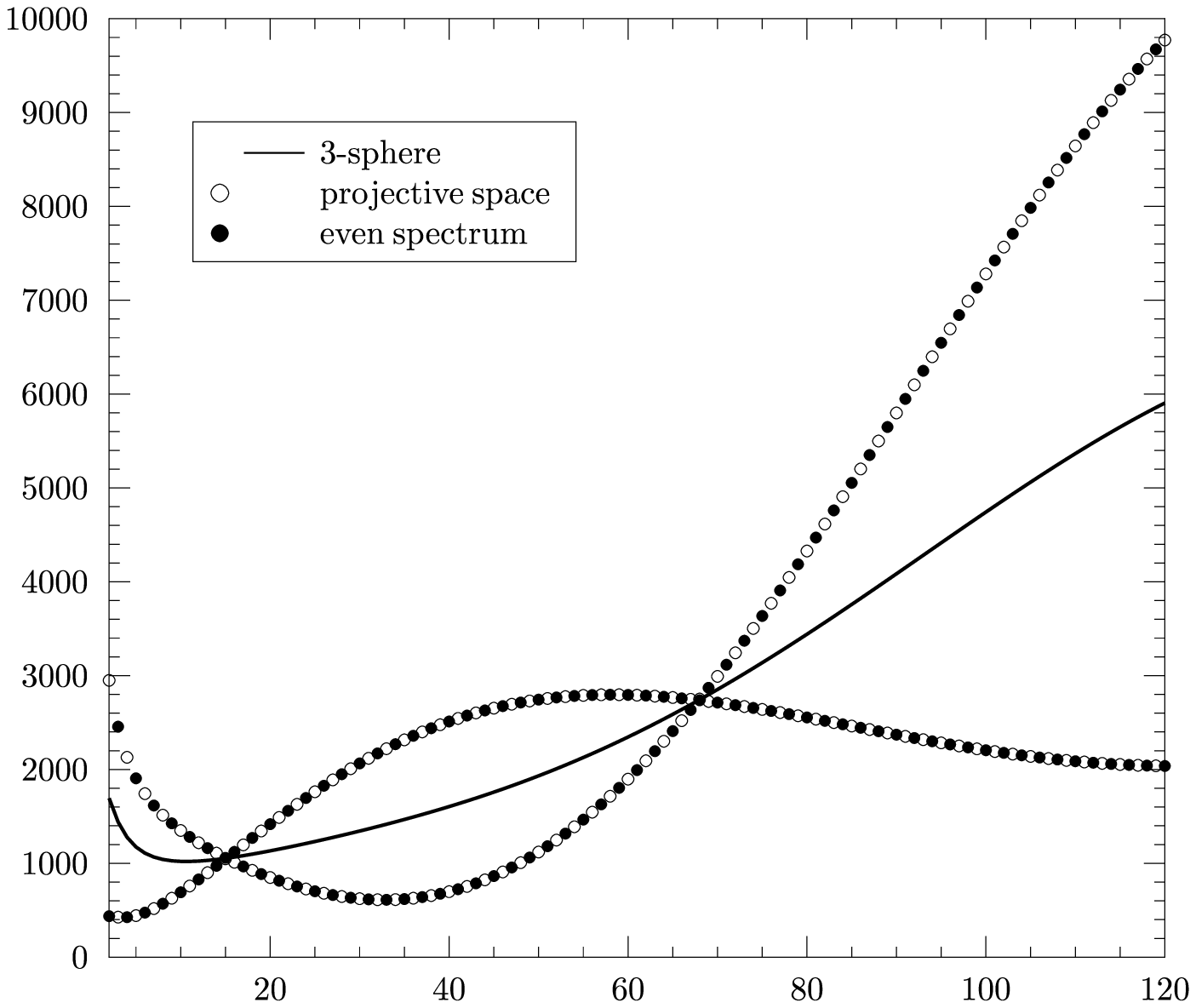}
\put(-262,190){$\delta T_{l}^{2}$}
\put(-55,20){$l$}
\put(-70,188){f)}
\vspace*{-35pt}
}
\end{minipage}
\end{minipage}
\caption{\label{fig:spectrum_P3}
{Panel a) shows the total contribution of the  angular power spectrum 
$\delta T_{l}^{2}$ of the projective space ${\mathbb P}^3$ calculated in the
tight coupling approximation for the cosmological parameters 
$\Omega_{\hbox{\scriptsize cdm}} = 0.238$, $\Omega_{\hbox{\scriptsize b}} = 0.0485$,
$\Omega_{\Lambda} = 0.9215$, $h=0.681$, and $n_{\hbox{\scriptsize s}}= 0.961$.
The outcome of these parameters is a distance to the surface of 
the last scattering $\tau_{\hbox{\scriptsize SLS}}\approx \pi/2$.
In panel b), c) and d) the corresponding Sachs-Wolfe, Doppler and integrated 
Sachs-Wolfe contributions are represented.  
The associated 2-point correlation functions $C(\vartheta)$} are shown
in panel e).
The angular power spectrum of the projective space ${\mathbb P}^3$,
the 3-sphere ${\mathbb S}^3$, and calculated from the even spectrum of $\beta$
are diagrammed in f).
}
\end{figure}

Let us now turn to a distinct feature visible in
figs.\,\ref{fig:S60}a and \ref{fig:S60}b.
At a total energy density $\Omega_{\hbox{\scriptsize tot}} = 1.208$,
a spike towards higher values of 
$\frac{S_{{\cal M}}(60^{\circ})}{S_{{\mathbb S}^3}(60^{\circ})}$
is present which occurs in all considered topologies.
This is due to the fact that for these cosmological parameters,
the distance $\tau_{\hbox{\scriptsize SLS}}$ of the observer from
the surface of last scattering is
$\tau_{\hbox{\scriptsize SLS}}\approx \pi/2$.
In this special case antipodal points on the sky having a distance
$\tau_{\hbox{\scriptsize SLS}} = \pi/2$ are identified
and have thus the same intrinsic temperature fluctuations.
Although this antipodal symmetry is not observed,
it is nevertheless worthwhile to discuss its implications.
This does not mean that the antipodal symmetry is perfectly mirrored
on the CMB sky,
since the CMB signal is composed of several contributions
with different properties.
To illuminate this point,
fig.\,\ref{fig:spectrum_P3} shows the main contributions separately
which are computed in the tight coupling approximation using
the cosmological parameters 
$\Omega_{\hbox{\scriptsize cdm}} = 0.238$, $\Omega_{\hbox{\scriptsize b}} = 0.0485$,
$\Omega_{\Lambda} = 0.9215$, $h=0.681$, and $n_{\hbox{\scriptsize s}}= 0.961$.
The computations are carried out for the projective space ${\mathbb P}^3$
which has the property that the multiplicities of even wave numbers
$\beta$ vanishes.
This is a common property of all spherical space forms except the
spherical manifolds which are determined by an group of deck transformation
with odd order.
For the antipodal symmetry, it is important to recognise from
the relation $Y_{lm}(-\hat n) = (-1)^l\,Y_{lm}(\hat n)$
that even and odd multipoles have to be considered separately.
The angular power spectra $\delta T_l^2$
shown in figs.\,\ref{fig:spectrum_P3}a-d display even multipoles as
open circles and odd multipoles as full disks.
Strong fluctuations of $\delta T_l^2$ occur in the total CMB signal
with respect to odd and even multipoles
as shown in fig.\,\ref{fig:spectrum_P3}a.
The usual Sachs-Wolfe contribution shown fig.\,\ref{fig:spectrum_P3}b
must vanish for odd multipoles because of the antipodal symmetry. 
The reverse behaviour occurs for the Doppler contribution
where the even multipoles must vanish, see fig.\,\ref{fig:spectrum_P3}c.
This is due to the vector character of the velocities
which arise as the derivative of a scalar velocity potential.
The third important contribution at large angular scales is the integrated
Sachs-Wolfe (ISW) contribution shown in fig.\,\ref{fig:spectrum_P3}d
which arises partly close to the surface of last scattering
(early ISW) and partly on the line of sight later on (late ISW).
The early ISW possesses an even-odd asymmetry as visible in
fig.\,\ref{fig:spectrum_P3}d.

The total angular power spectrum $\delta T_l^2$ shown in
fig.\,\ref{fig:spectrum_P3}a is composed of these three contributions
at large angular scales.
For $l < 15$ the usual Sachs-Wolfe contribution (fig.\,\ref{fig:spectrum_P3}b)
is the most important contribution,
and thus the asymmetry of the total $\delta T_l^2$ is of the kind
that even multipoles $l$ dominate the odd ones.
In the range $15<l<70$ the Doppler contribution dominates
leading to a reversal of the even-odd asymmetry in the total $\delta T_l^2$,
see fig.\,\ref{fig:spectrum_P3}a.
For $l>70$ the usual Sachs-Wolfe contribution dominates again
leading to a further reversal of the even-odd asymmetry.
In this way the complex behaviour of the angular power spectrum $\delta T_l^2$
can be understood.
Although this discussion is restricted to the special case
$\tau_{\hbox{\scriptsize SLS}} = \pi/2$,
the numeric shows that cosmological models having
values of $\tau_{\hbox{\scriptsize SLS}}$ not too far from $\pi/2$
display such a behaviour.
So this asymmetric behaviour survives for a large class of models.

A comparison of the projective space ${\mathbb P}^3$ with
the simply-connected ${\mathbb S}^3$ can be found
in fig.\,\ref{fig:spectrum_P3}f.
Since the simply-connected ${\mathbb S}^3$ has odd and even wave numbers,
no even-odd asymmetry occurs in this space form.
The projective space ${\mathbb P}^3$ possesses this asymmetry as discussed
due to its odd wave numbers.
Although there exists no spherical space form
which has only even wave numbers,
it is interesting to compute the angular power spectrum $\delta T_l^2$
for an artificial spectrum having only even wave numbers.
The result is shown in fig.\,\ref{fig:spectrum_P3}f, and
one observes its complementary behaviour with respect to
the projective space ${\mathbb P}^3$.
This neatly explains why no even-odd asymmetry occurs
in the case of the ${\mathbb S}^3$ space.

The even-odd asymmetry determines the behaviour of
the correlation function $C(\vartheta)$ at angles
around $\vartheta = 180^\circ$.
In fig.\,\ref{fig:spectrum_P3}e the different contributions are
shown separately.
The asymmetry of the usual Sachs-Wolfe contribution leads to
large positive values at $\vartheta = 180^\circ$,
whereas the Doppler contribution gives large negative values.
This behaviour follows from eq.\,(\ref{Eq:C_theta_Cl})
and $P_l(-1) = (-1)^l$.
Due to the dominance of the usual Sachs-Wolfe contribution
at large scales, the total correlation function $C(\vartheta)$
possesses large values for $\vartheta \gtrsim 140^\circ$.
Since a low power is observed in the CMB sky at these scales,
models with $\tau_{\hbox{\scriptsize SLS}} \simeq \pi/2$ are unrealistic.


\section{Summary and Discussion}

In this paper we study spherical models of our Universe
which possess spatial spaces that are multi-connected.
These spaces with positive spatial curvature arise
by tiling the 3-sphere ${\mathbb S}^3$ using a deck group $H$.
We restrict attention to groups of order 8
such that the volume of the considered manifolds is
$\hbox{vol}({\mathbb S}^3)/8 = 2\pi^2/8$.
There are two such lens spaces $L(8,1)$ and $L(8,3)$,
and the $D_8^*$ manifold
which is obtained by the binary dihedral group $D_8^*$.
Furthermore, there are two cubic Platonic manifolds $N2$ and $N3$,
however, it is shown that $N2$ and $L(8,3)$ are equivalent
and $N3$ corresponds to $D_8^*$.

Such a multi-connected space can be represented by its fundamental domain
which is the subset of points such that no element of 
$g \in H,\: g\neq e$ can operate  inside the domain, 
but any point of the cover outside the domain can be reached 
by the action of $H$ on a point inside the domain.
Such a cell is called a Voronoi domain.
It is the natural domain for an observer sitting at the origin
of the coordinate system for which the group elements $g\in H$ are defined.
If the group elements $g\in H$ are independent of the choice of
the position of the observer,
all observers construct the same Voronoi domain.
Such manifolds are called homogeneous.
On the other hand, if there is such a dependence,
the shape of the Voronoi domain varies and such manifolds
are called inhomogeneous.
From the above manifolds, only $L(8,3) \equiv N2$ is inhomogeneous.
One can construct a spatial curve along which the observer can be shifted
such that at one position the lens shaped Voronoi domain emerges
whereas at an other position,
the cubic Platonic Voronoi domain $N2$ appears.
Other positions do not lead to domains having special properties.

The important point for cosmic topology is
that the statistical properties of the CMB anisotropies depend on the
group elements $g\in H$
defined for an observer that sits at the origin of the
coordinate system.
Thus, inhomogeneous manifolds allow a much richer variety of CMB anisotropies,
and the comparison with observations is much more involved.
For the multi-connected spherical spaces generated by groups $H$ of order 8,
the temperature 2-point correlation function $C(\vartheta)$,
eq.\,(\ref{Eq:C_theta}), and
the multipole moments $C_l$, eq.\,(\ref{Eq:Cl_ensemble_general}),
are computed for the CMB anisotropies.
A suitable measure for the suppression of the anisotropies at large
angular scales is the $S(60^{\circ})$ statistics (\ref{Eq:S_60})
which emphasises the unusual behaviour at scales $\vartheta\geq 60^{\circ}$.
The focus is put on the inhomogeneous space $L(8,3) \equiv N2$,
for which a one dimensional sequence of observer positions is derived
which exhaust all possibilities allowed by this inhomogeneous space.
The CMB anisotropies are calculated for this sequence, and it turns out
that the strongest large scale suppression occurs for the
Platonic cubic Voronoi domain, which has the observer in the centre
of the $N2$ cell.
On the other hand, the least suppression is observed for the lens shaped domain.
This is in agreement with the hypothesis
that well proportioned domains give the largest suppression of power.
Thus, the example of the inhomogeneous lens space
$L(8,3)\equiv N2$ demonstrates 
that it can be premature to classify all lens spaces as
not-well proportioned domains and thus uninteresting with respect to
the observed CMB power suppression.
Nevertheless, for $\Omega_{\hbox{\scriptsize tot}}>1.07$,
the homogeneous Platonic space $N3$ leads to an
even stronger suppression of large scale CMB anisotropy
than it is the case for all observer positions in $N2$.

The special case that appears
when the distance $\tau_{\hbox{\scriptsize SLS}}$
to the surface of last scattering satisfies
$\tau_{\hbox{\scriptsize SLS}} = \pi/2$ is studied,
although the cosmological parameters are unrealistic in this case.
Then the topology requires an antipodal symmetry
which partly survives in the CMB anisotropies and is reflected
in the $S(60^{\circ})$ statistics.
It is discussed in detail how the different contributions affect the
CMB signal in distinct ways as presented in fig.\,\ref{fig:spectrum_P3}.


\section*{Acknowledgements}

We would like to thank the Deutsche Forschungsgemeinschaft
for financial support (AU 169/1-1).
The WMAP data from the LAMBDA website (lambda.gsfc.nasa.gov)
were used in this work.


\end{document}